\documentclass[AMA,STIX1COL]{WileyNJD-v2}

\usepackage{booktabs}
\usepackage[textsize=scriptsize,textwidth=1cm]{todonotes}

\articletype{Article Type}%

\received{}
\revised{}
\accepted{}
        
\begin{document}

\title{Emotional Attachment Framework for People-Oriented Software}

\author[1]{Mohammadhossein Sherkat*}
\author[1]{Antonette Mendoza}
\author[1]{Tim Miller}
\author[1]{Rachel Burrows}

\authormark{SHERKAT \textsc{et al}}

\address[]{\orgdiv{School of Computing and Information Systems}, \orgname{The University of Melbourne}, \orgaddress{\state{VIC}, \country{Australia}}}

\corres{*Mohammadhossein Sherkat, School of Computing and Information Systems, The University of Melbourne, VIC 3010, Australia. \email{sherkatm@unimelb.edu.au}}


\abstract[Summary]{In organizational and commercial settings, people often have clear roles and workflows against which functional and non-functional requirements can be extracted. However, in more social settings, such as platforms for enhancing social interaction, successful applications are driven by user emotional engagement than functionality, and the drivers of user engagement are difficult to identify.
A key challenge is to capture and understand users emotional requirements so that they can be incorporated into interaction design.

This paper proposes a novel framework called the \emph {Emotional Attachment Framework} (EAF), which is based on existing models and theories of emotional attachment 
to facilitate the process of capturing and understanding emotional goals in software design.
To demonstrate the method in use, a two-part evaluation is conducted. First, emotional goals are elicited for a software application that aims to provide help for homeless people in Australia. The results are evaluated by domain experts and compared with an alternative approach. The analysis shows that the root of majority of homeless people's emotional goals goes back to their needs for expressing personal identity as well as seeking pleasure. Second, we ran a semi-controlled experiment in which 12 participants were asked to apply EAF and another techniques on part of our case study. The outcomes show that EAF has the potential to help system analysts uncover additional emotional requirements as well as provide valuable insights into these emotional goals.


}

\keywords{People-Oriented Software, Requirements Engineering, Emotional Goals, Emotional Attachment Framework}

\maketitle

\section{Introduction}
\label{sec:intro}
\begin{itquote}
Computing is not about computers any more. It is about living. --- Nicholas Negroponte  \cite[p.\ 197]{calvo2014positive}.
\end{itquote}


The main goal of Requirements Engineering (RE) is to understand what people want and desire, so that we can design and implement a successful software system \cite{guinan1986development,dieste2008understanding,gonzales2011eliciting,colomo2010study}.
In this process, stakeholders must decide on the most effective combination of software features. To make this decision, two critical activities are requirements elicitation and requirements modelling.
This process has been described as ``getting into someone's head'' to capture his or her needs and desires \cite {guinan1986development}. 

Despite the maturity of existing requirements elicitation processes, many software applications struggle to engage people
\cite {clancy1995standish, OASIG1995, whittaker1999went, tichy2008business, gonzales2011eliciting}, ultimately leading to people rejecting the software \cite {van2001interactive, platt2007software, Dix:2003:HI:1203012, 1663532}.
It is widely argued that a major concern is that software engineers seem to only focus on traditional functional and quality requirements and overlook key drivers of engagement such as people's emotions and values \cite {bentley2002putting, draper1999analysing, gogueny1994requirements,hassenzahl2001engineering,krumbholz2000implementing,miller2015emotion, proynova2011investigating}. It is therefore important to take into account software requirements that will create the desire to engage with the system --- in addition to functional and quality requirements. These requirements may be related to social values or emotions that people want to feel. In this research, we call such requirements  \emph{emotional goals} \cite {sutcliffe2010analysing, miller2012understanding}.

Addressing emotional goals in design is not straightforward and software engineers face several barriers due to the nature of emotions. First, identifying and understanding people's emotional goals is difficult as they are the subjective attributes of people and not the property of software \cite{callele2006emotional,mendoza2013role}. People rarely are able to express their emotions directly. Although people recognize their emotions when they encounter them, it is difficult for them to articulate them beforehand since emotions are a subjective part of their consciousness \cite{Goguen:1994:RER:177970.184582}. Second, even if people are able to express emotional goals, there is limited knowledge about how these emotions can be captured and then translated into concrete requirements as part of design \cite{salzer1999atrs}. Third, the unstructured and ambiguous nature of emotional goals make converting them into a software specification difficult. Thus, the lack of a systematic methodology to map such intangible emotions to a tangible set software features remain a challenge. 
Consequences related to overlooked emotional goals 
are particularly profound in social applications
such as platforms for enhancing social interaction, social networking and public health software systems. In these types of software systems users posses a variety of characteristics, including their needs, goals, motivations, and lifestyle that ultimately influence the ability of software to appeal and engage. 
In this paper we define these types of software systems as \emph{{P}eople-{O}riented {S}oftware} {(POS)} systems. In POS systems, requirements elicitation is challenging as people are often not obliged to use a software system, do not generally have well-defined roles and responsibilities, have different cultural and social backgrounds, and do not necessarily have the same requirements as each other \cite{miller2015emotion}.
%
Traditional requirements engineering methods capitalise on well-defined business work-flows and tasks that are less applicable in contexts that are not as structured or protocol-driven like emotional goals in POS. 

In terms of scientific theories, there are numerous theories and frameworks that have categorized the human's emotion, such as Parrott's Emotions by Groups \cite{parrott2001emotions}, Plutchik's Wheel of Emotions \cite{plutchik2003emotions}, Ekman's Six Basic Emotions \cite {ekman2007emotions, lama2008emotional} and the Hourglass of Emotions \cite{ cambria2012hourglass}. 
While some ideas from these theories can be used for considering emotional goals in software engineering, they often are not suitable for the design of software systems for several reasons: (i) they typically assume emotions operate independently of users' cognitive processes; (ii) they do not model the fact that people may experience two or more emotions at the same, such as feeling independent or connected; (iii) they cannot be used to describe the complex range of emotions that people may feel in daily use of a software system; and (iv) are not contextualized in everyday life (i.e. they are too abstract) \cite{cambria2012hourglass, smith2009critiquing}. These limitations motivate the need for bridging the gap between the psychological frameworks and software engineering. Accordingly, a key question at the heart of successful requirements engineering activities is: \emph{How can we get a better understanding of people's emotional goals in designing a software system?}

In this paper we investigate the nature of emotional goals within the context of requirements engineering. 
Our research question is: 
\emph {How can system analysts achieve a better perception regarding stakeholders' emotional goals in the process of requirements engineering?} 
We present a novel framework called the \emph{Emotional Attachment Framework}, which is mainly based on theories and models in psychology \cite{arnold1970perennial, Desmet2002}.

In this research, we do not propose a new taxonomy of emotional categories as we already have them in the field of psychology \cite {cornelius1996science, ekman2007emotions, lama2008emotional, thoits1989sociology, sep-emotions-17th18th, arnold1970perennial, Desmet2002, cambria2012hourglass}. Instead, we provide a framework to increase our insights in capturing and
understanding emotional goals. This is based on existing models, theories and taxonomies of emotions in psychology. We also outline a process model for using the proposal framework in practice.
Understanding emotional goals is an essential precondition for detailed design and development efforts, therefore the focus of this work spans this first step of gaining deeper insights in capturing and understanding emotional goals.



We evaluate our proposed framework in two parts.
First, using an industry case study --- a mobile website to help homeless people find information about available support services. 
We used interview data at hand to analyze and extract emotional goals. We then compared our findings with an alternative approach for eliciting emotional goals \cite{miller2015emotion}. The comparison analysis with the alternative emotional goals elicitation approach indicated that we discovered more subtle emotional goals by using our proposed framework.
Further, we tested the applicability and effectiveness of our proposed framework by interviewing four independent domain experts involved in this case study. The results show that our approach identified and categorized people's emotional goals that the domain experts agreed with, and that they believe the elicited emotional goals provide improved support for requirements engineering.

Second, we recruited $12$ participants with software engineering background and asked them to complete a series of tasks and answer a series of questions about EAF and alternative analysis technique. We measured the time and accuracy of their responses and then asked a set of qualitative questions around their preferences between the techniques. We then asked domain experts to analyze the participants' results and evaluate their output.
The results show that the EAF is more time efficient, easier to use and learn, and helped participants generate more non-repetitive and relevant insights into emotional requirements of the system.




The following section presents related literature in the area  emotional goals and prior efforts in considering emotional goals in design. In Section~\ref{sec:proposedmethod}, we present our framework for understanding the people's emotional goals and this proposed framework will be evaluated in Sections~\ref{sec:evaluation}. The last two sections of this paper are dedicated to discussion and conclusion.

\section{Literature review and related work}
\label{sec:background}


Frog Design founder Esslinger states that:  {``even if a design is elegant and functional, it will not have a place in our lives unless it can appeal at a deeper level, to our emotions"} (cited in \cite [p.9]{sweet1999frog}).
In this section we briefly review the nature and foundations of emotions and what emotional goals are. We also review the efforts in considering emotional requirements in design and development including software system domain.

\subsection {Concepts and Definitions}

The importance of considering emotions is recognized in software system design, however, this subject has been of interest in different fields of science and research such as philosophy \cite{sep-emotions-17th18th}, biology \cite{de1990rationality}, physiology  \cite {de1990rationality}, sociology \cite {thoits1989sociology} and psychology \cite{Desmet2002}. It means that the main roles and concepts of emotions have been borrowed by the field of software system design from other domains. In this section 
it is important to recognize the characteristics of emotional goals and the aspects which make them distinguish them from other types of requirements. In order to understand emotional goals, it is first essential to have an understanding of the concept of emotion.

\subsubsection{Definition of Emotion} 
\label{emotionconcept}

Although it seems the definition of emotion is clear, as Arnold \cite {arnold1970perennial} mentioned, understanding the nature of emotions, especially emotions' characteristics, how do emotions work?, how are emotions felt and expressed?, etc. is a lasting problem. Emotions influence everything we do from the way we behave and think, to the way we communicate and make decisions. People are capable of a vast range of emotions, from satisfaction in daily tasks to the sadness of losing valuable things and death of a loved one \cite {ortony1990cognitive, norman2005emotional}.

By reviewing the literature, we found several different terms for reflecting people's different feelings. Affect, emotion, mood, passions, value, motivation and sentiment are just a few to name. Although the meaning of these words differ from one another and refer to different concepts, they are routinely used interchangeably. Among all of these words, the word {``emotion"} is often used in a wide variety of research domains. In this research we also use the term emotion, as this is a common term that people use to refer their subjective and hard-to-measure feelings and it is more related to people rather than software \cite{miller2015emotion}. 
In this research what we mean by the people's emotions is the people's different feelings and mental states which have different characteristics and effects on how people make decisions and behave.

\subsubsection{Forming Emotion} 
\label{formingemotion}

Prior studies in the field of product design have tried to understand how consumers' emotions form  \cite {Desmet2002}. 
As Arnold, the founder of `Appraisal Theory' argued, people form emotions when encountering a product or service. This is the consequence of their assessment (or their \emph{`appraisal'}) of the product or service capabilities and how it may harm or benefit them\cite {arnold1970perennial}. 
Based on the Arnold's definition, an appraisal is the \emph{``direct, immediate sense of judgment of weal and woe"} (cited in \cite [p.9]{cowie2001emotion}). He also defined emotion as the \emph{``felt tendency toward anything intuitively appraised as good or away from anything  intuitively appraised as bad"} (cited in \cite [p.33]{rulla2003depth}). Based on these definitions without appraisal there is no emotion and positive or negative emotions are the consequence of appraising a product or a service beneficial or harmful respectively. 
In other words, people can experience different emotions such as joy, frustration and desire in response to products that they appraise as matching or mismatching with different emotional goals that they have. 

Desmet and Hekkert argued that what causes humans' emotions is their interpretation of a product or a service rather than the product or service itself \cite{desmet2007framework}. This interpretation is the consequence of cognitive appraisal process which often happens automatically and unconsciously in the people's minds. Desmet used Arnold's appraisal theory to visualize and develop this concept as the `Basic Model of Emotions'. Desmet's model contains three variables: Appraisal, Concerns (goals) and Stimulus (product or service) (Figure~\ref{fig:fig1}) \cite{Desmet2002}. The definitions of these three variables are as below:

\begin{figure}[ht]
\centering
\includegraphics [scale=0.7]{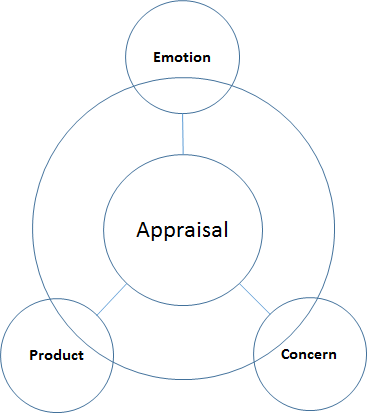}
\caption{Desmet's Basic Model of Emotions \cite{Desmet2002}}
\label{fig:fig1}
\end{figure}

\begin{description}
    \item [Appraisal] According to appraisal theory, appraisal refers to automatic and unconscious evaluation of a product or service for the people's personal concerns (goals) \cite{scherer2001appraisal}. The output of this evaluation in Desmet's basic model of emotions shapes the people's positive or negative emotions. 
    \item[Concerns] Based on what Frijda has discussed, behind each emotion there is a concern \cite{Frijda2010handbook}. If people's concerns can be addressed by product or service capabilities, the product or service is appraised as beneficial and lead to positive emotions. In the opposite side, the product or service capabilities that cannot match with the people's concerns are appraised as harmful and lead to negative emotions. Desmet \cite{Desmet2002} categorized concerns in two main groups, universal (such as the concern for safety and love) and culture and context-dependent (personal concerns such as concern for being home late or missing good seat at the cinema).
   
    \item [Stimulus] In Desmet's basic model of emotions, the stimulus refers to the element labelled as `Product` in Figure~\ref{fig:fig1}. It represents any product or service that can address one or more concerns. 
    Based on the Desmet's definition, physical products, services, remembered and imaginary items can be considered in the basic model of emotions as a stimulus. 
\end{description}

\subsubsection{Emotional Attachment}
\label{sec:emattach}
Emotional attachment is an attempt to create emotional links between people and products or services. It has received a lot of attention in recent years particularly in the field of marketing \cite{11015023620150701, norman2013design, guo2014emotional}. 
The main basis of emotional attachment is derived from `Attachment Theory' was initially developed by Bowlby to explain the relationship between the parents and infants \cite{bretherton1992origins}. 
However in last two decades a wider range of relationships and contexts have been encompassed. Recent studies indicate that individuals not only emotional attach to other individuals, but also to brands \cite{whan2010brand, thomson2005ties},
pets \cite {sable1995pets}, gifts \cite {mick1990self}, places \cite{giuliani2003theory, scannell2010defining}, collectables \cite{slater2001collecting}, firms and organizations \cite {mende2011attachment} and, objects \cite {ainsworth1969object, page2014product}.

People experience emotional attachment at different levels of abstraction. They may emotionally attach to a product or a service in general or specific capabilities. For instance, people may emotionally attach to Facebook\texttrademark\ (as a general social network application) or like one or more capability(ies) (sharing media as a specific capability).


Mugge \cite{Mugge} proposed four categories of sources for creating product emotional attachment:

\begin{description}

\item [Self-expression] 
People desire a distinct personal identity from others, what Mugge et al., call  \emph{Self-expression} \cite {mugge2008product}. Based on Collins on-line dictionary, self-expression refers to the expression of personal identity such as feelings, thoughts or ideas\footnote{http://www.collinslanguage.com}. Past studies argue that people become more attached to products or services if they can show their personal identity \cite {ball1992role, kleine1995possession}. 

Further, Mugge et al. \cite {mugge2009emotional, mugge2009development} show that more self-expressive occurs if people can `embed their personality' into a product. Embedding the personality is based on the `self-congruity theory', which says humans need to express and create a positive and consistent view of one self \cite {govers2005product}. Accordingly people attach emotionally to a product or services that are reflect their own `self-concept'. 

Sirgy \cite{sirgy1982self} shows that this `self-concept' has two categories: (1) (\textbf{\emph{Ideal Self}}), which is the view people have of themselves; and (2) (\textbf{\emph{Public Self}}), which is the image people would like other people to have of them.

\item [Affiliation]
The role of affiliation in forming emotional attachments stems from the people's desires to have relationships, to be connected, associated, and involved with others \cite{meschtscherjakov2014mobile}. Murray's Theory of Human Personality describes people wanting to feel a sense of affiliation with others and belonging to social groups \cite{champoux2016organizational}. 
Although it may seem that the affiliation opposes self-expression, they can exist simultaneously. Kleine et al.\  argued that although people like to establish a unique identity, they are also motivated to have interpersonal connections \cite{kleine1995possession}. A good example of coexisting affiliation and self-expression is Facebook\texttrademark. A Facebook\texttrademark\ page expresses a unique identity by writing posts and sharing individual pictures (self-expression) and simultaneously he/she is a member of specific groups (affiliation).

\item [Memories]  
Kleine notes that people like to maintain a sense of the past, happy moments in life, and be reminded of people, occasions, places etc.\ that are important for them \cite {klein2012memory}. 
Csikszentmihalyi and Halton argued that if a product has clues or symbols for their consumers regarding ``where they have been, who they are now and what they aspire to be'', these symbols can attached them emotionally to the product \cite{csikszentmihalyi1981meaning}.
Memories can be both positive and negative. Accordingly, people like to remember or re-enforce positive memories and forget negative ones. 
Mugge contends that if people experience positive memories in using a product, they emotionally attach to it regardless of its utility and appearance \cite {mugge2010product}.

\item [Pleasure] 
Based on the Hedonism school of thought, pleasure and happiness are the primary aim of human life \cite{Stanford}. Accordingly, if a product provides sensory pleasure and enjoyable activity, it causes emotional attachment \cite {Mugge}. 
Jordan defines pleasure in designing a product as ``emotional, hedonic and practical benefits associated with products'' \cite [p.11]{jordan2002designing}
which appears in different forms including (i) \emph{\textbf{Physical Pleasure:}} relates to people physical body and the pleasure experienced by sensory perception - five senses, (ii) \emph{\textbf{Social Pleasure:}} relates to people relationships with others and with status and image, and (iii) \emph{\textbf{Ideological Pleasure:}} relates to peoples values and beliefs \cite{jordan2002designing}.


\end{description}

It seems reasonable to extend the boundary of emotional attachment and assume that people can attach emotionally to a software system as a type of product or service.
Based on the emotional attachment definitions in other disciplines we define this concept in software engineering as \emph {`the strength of the emotional relationship people experience with a software system'}. People's emotional attachment will determine whether people use the software system as a beloved tool, they simply put up with, reject it, or never appropriate it at all.

\subsubsection{Emotional Goals}



Based on the definitions of emotion, emotional goals can be defined as the people's psychological or mental goals which have intrapsychic origin \cite {glanze1990mosby}. Callele et al., define emotional goals as a way to help system analysts to deliver emotional experience to people  in order to achieve optimal software system experience \cite{callele2008balancing}.
Lopez-Lorca et al., have considered emotional goals in two different categories. 1) Personal Emotional Goals which represent how people want to feel as a general sense and independent of any particular system (feeling independent) and, 2) System-dependent Emotional Goals which represent people's desires towards a specific software system \cite{lopez2014modelling}.

In this research we define emotional goals as \emph{what people would or would not like to feel by using a software system}.  As the root of functional or non functional capabilities goes back to people's functional or non functional requirements, we can define people's emotional goals as what people would or would not like to feel in the process of satisfying their functional or non functional requirements. Recent research has demonstrated how emotional goals are distinct from the application's non-functional requirements\cite{miller2015emotion}. If we consider the social application of Facebook\texttrademark\, it is possible to see how a functional requirement of connecting friends is quite different from the emotional goal of feeling connected. Alternatively, we can see how the quality goal of being secure is also quite different to feeling secure from the perspective of the user.

Emotional goals are influenced by people's values, aspirations, socio-cultural norms and mood that people would like to experience when using a particular software system. The interplay of these social variables further reinforces the idea that emotional goals cannot be confined to immediate sensory pleasures, but are in fact more reflective and long-term.  


\subsection{Emotional Goals in Design and Software Engineering}
Research into emotional goals in design dates back to the 1970s. Most of the research in this area focuses more generally on product design \cite{jordan2002designing, desmet2007framework, norman2005emotional, hassenzahl2001engineering, chapman2015emotionally, schifferstein2004designing}. 
Although emotional goals are not emphasised in software engineering, there is research in the field of human-computer interaction \cite {ingram1984designing, marcus2015emotion} and emotionally intelligent software agents \cite {bates1994role}. 
Considering emotional goals is imperative in game design \cite {norman2013design,salen2004rules} where the main goal of such games is to `have fun' \cite {callele2006emotional}. According to Draper \cite{draper1999analysing}, fun is a dependency between the game application and players' goals and as such is not a property of game application itself. 



Ramos and Berry \cite {ramos2005emotion} investigated software features that caused fear (as a type of emotion) to show the influence of emotion in system acceptance. They analyzed an application that stores information about mistakes within a company, and who was responsible for them to show the relationship between negative emotions and system acceptance. Although in their value taxonomy they categorized people's soft goals in four categories including structural, social, political and symbolic, however, they did not propose specific guideline for eliciting people's emotion. 

Colomo-Palacios et al. \cite {colomo2011using} developed a method called the `Affect Grid' to understand the effect of  emotions on software requirements. 
Their results indicated that emotions were a factor to take into account in establishing requirements stability. Colomo-Palacios et al. concluded that understanding the stakeholders' emotions includes knowing ``the reliability and stability of the definition of those requirements" \cite {colomo2011using}.

Hassenzahl et al., \cite{hassenzahl2001engineering} considered the people's emotional goals as a `hedonic quality'. In their study, hedonic quality is an aspect of a software system (especially in graphical user interface) that creates positive subjective experience in the people based on their emotions. According to Hassenzahl et al., hedonic qualities are non-task-oriented qualities which prevent product or service consumers from boredom and discomfort, and create motivation, stimulation and challenge. 
Thew and Sutcliffe \cite{thewvalue, thew2008investigating} in their studies reviewed different issues related to poor understanding of stakeholder values and proposed a taxonomy of stakeholders` values, motivations and emotions (VME). 
In their taxonomy they suggested eight, nine and six upper level categories for value, motivation and emotions respectively. Although they proposed the elicitation guides in the form of conversation topics for each category, the proposed taxonomy makes no recommendations for solving value clashes.

Probably the most famous studies in considering the people's emotions in designing products can be seen in Norman's studies \cite{norman2005emotional}. Norman believes that there are three levels of emotion:  visceral, behavioural and reflective. The visceral level is pre-conscious formed automatically and related to product appearance such as colours and style. 
The behavioural level is sub-conscious, and related to the use and experience with a product based on function, performance, and usability. The reflective level is conscious, and as Norman stated ``the highest level of feelings, emotions and cognition reside" in this level.  Accordingly the visceral and behavioural levels are about the ``now", while the reflective level is longer over time and is more about the ``satisfaction" produced by using a product and main causes of long-term attachments. Miller et al.\ \cite{miller2015emotion} argue that the reflective level has been neglected in software engineering research and practice. As they have argued, many social goals like "fun" are complex and reflective and not immediately measurable.

\subsection{Eliciting Emotional Goals}

Although there are limited studies about how to deal with people's emotional goals in software engineering, perhaps the initial efforts in highlighting the importance of eliciting and capturing people's emotional requirements was escalated in Mumford's ETHICS method  \cite{mumford2013values}. Although this method encouraged system analysts to find ways to capture people's values and emotions in system design process, it is a questionnaire-based technique with less flexibility and people may not be comfortable with this time-consuming method.

Another technique that has been applied widely for eliciting and capturing people's soft goals is ethnography \cite {sommerville1993integrating, hughes1995presenting, viller1999social}. In this technique, a system analyst gather data including observing people in their workplace and may participate in their everyday work to realize people's soft goals. Although this technique does not require asking questions directly for capturing people's soft goals, the quality of output depends entirely on ethnographer expertise as it does not have any guideline supporting ethnographers in soft goals elicitation process. Yu \cite {eric2009social, eric2010social} by proposing the \emph{i*} modelling notation tried to model \emph{soft goals} such as as ``trustworthy'', ``flexible'', ``minimal intrusion'' or ``normal lifestyle". However, the primary contribution of \emph{i*} accredited to its capability as a modelling language as opposed to an elicitation technique.

In the field of human-computer interaction, Friedman suggested Value-sensitive design  approach for eliciting human values \cite {friedman1996value, friedman2002value}. In this technique user scenarios and storyboards are used for eliciting people' feelings toward a developed software system or its prototype in the form of cue cards. In other words, this technique does not propose any guideline for eliciting people's soft goals such as values and emotions in the requirements engineering process. Friedman's idea was followed by Cockton et al. \cite{cockton2009evolving} in developing the Worth Maps technique for capturing  informal descriptions of people's feelings and emotions. 
In some other studies such as \cite{gordijn2003value, komssi2011integrating, fuentes2010understanding} instead of outlining a holistic and repeatable technique, only general advice have been given for eliciting specific types of people's feelings and emotions. 

Based on what we discussed in this section, we can argue that  although many of the previous methods have been successfully applied for requirements elicitation, they did not propose a systematic and repeatable process for capturing and understanding people's emotional goals for software development purposes. The current paper aims to address the above mentioned issues and provide something specific for POS.

\section{Proposed framework
}
\label{sec:proposedmethod}

To answer the research question in this study we first need to answer the question \emph{`How do people's emotions form during their encounter with a software system?'}. Understanding the process of forming emotions provides insight about the aspects that we need to investigate to realize emotional goals.  
In this section, we first present our conceptualisation of forming people's emotions, which we call \emph{Emotional Appraisal Model}. We then introduce our proposed framework -- the \emph{Emotional Attachment Framework} (EAF) -- and process model for understanding the people's emotional goals.


\subsection{Emotional Attachment Framework}
\label{sec:ETF}

According to the Basic Model of Emotion (Section \ref{formingemotion}), we can argue that forming people's emotions is the consequence of people's appraisal regarding whether a software system's capabilities satisfies their goals or not. If the capabilities address their goals, the software system is appraised as beneficial. Otherwise as unhelpful. Therefore, forming emotions is the outcome of match and mismatch of people's goals with software systems capabilities that result in a positive emotion, a negative emotion or neutral emotion.

\begin{figure}[ht]
\small
\centering
\includegraphics [scale=0.7]{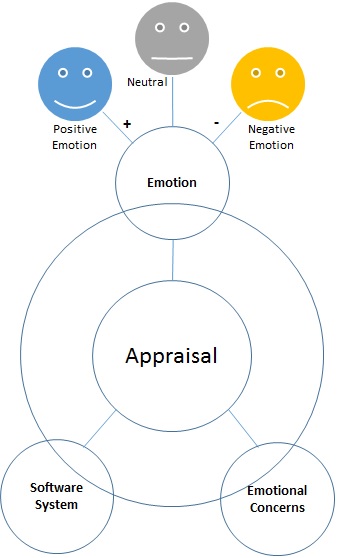}
\caption{Customized version of Desmet's Basic Model of Emotions}
\label{fig:fig2}
\end{figure}

We adapt and customize Desmet's model of emotions for our purpose 
in the software engineering  domain (Figure \ref{fig:fig2}), focusing on software applications instead of products more generally. In addition, Desmet's definition of a `concern' refers to all types of customer concerns (or goals), however, we adapt it to focus specifically on emotional goals. 



Figure \ref{fig:fig2} represents the relationship between emotions, emotional goals and a software system. The various types of positive, negative or neutral emotion are the result of match and mismatch between people's emotional goals with software system capabilities. It is clear that the type and importance of emotional goals depends both on people and the software application domain. The interplay between these aspects also varies over time. For instance in computer games, pleasure is prioritized, whereas in social network applications group affiliation could be more important.

Based on Figure \ref{fig:fig2}, forming an emotional attachment or detachment with a system depends on people's appraisal of their emotional goals and the system's capabilities. Accordingly, we can categorize the main drivers of forming emotional attachments under emotional concerns. Adding the \emph{drivers} of forming emotional attachment to the basic model forms a new framework that we call the \emph{Emotional Attachment Framework} (EAF). 

\begin{figure}
\small
\centering
\includegraphics [scale=0.5]{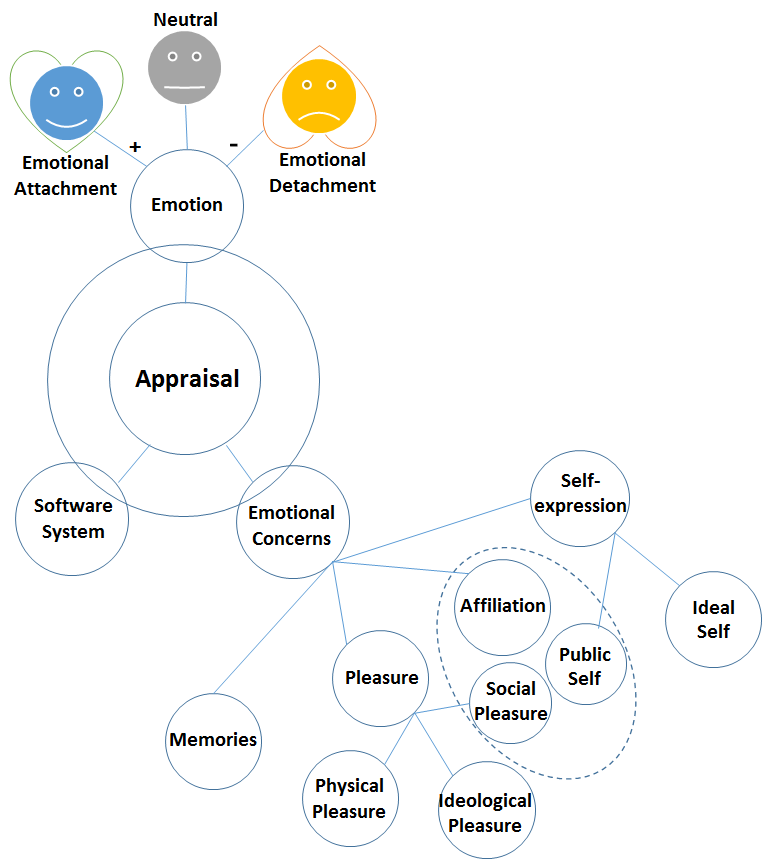}
\caption{Emotional Attachment Framework}
\label{fig:fig3}
\end{figure}

Figure \ref{fig:fig3} represents the schematic view of the EAF.
It is our assumption that although people may have different emotional goals, the root of their emotional goals can always be found by reinforcing what causes people to emotionally attach to (or detach from) a software system. Identifying these will support a better understanding of emotional goals in order to create the desire to engage with the system.
This perspective stems from this fact that people need to support their emotional sense and define and maintain their self-concept (i.e. beliefs about themselves) \cite{ball1992role, kleine2004integrative}.
For this purpose, we have adopted the core ideas of customer emotional attachment from the field of marketing and used Mugge's attachment drivers \cite {mugge2008product, mugge2009emotional, mugge2009development, mugge2010product, Mugge} (discussed in Section \ref{sec:emattach}).
The main reason we use Mugge's model is that it offers the most complete mode of emotional attachment in the existing literature.

As Figure \ref{fig:fig3} represents, according to the four sources of creating emotional attachment in Mugge's study, people's emotional goals are driven by their aspirations related to: (i) self-expression; (ii) affiliation; (iii) pleasure; and (iv) memories.



\paragraph{Self-expression}

As we reviewed Mugge et al., arguments in Section \ref{sec:emattach}, people become emotionally attached to a product if it can help them to display their individuality (e.g. status among peers) \cite{mugge2008product, mugge2009emotional, mugge2009development}. This can be extended to software products. For instance, a person may experience emotional attachment with a software system for educating society about climate change as it expresses their identity of being an environmentally conscious person \cite{winfree2017learning}. 
As an other example, in an emergency system for older people \cite{miller2015emotion}, the sense of `independence' or `empowerment' facilitates self-expression between people.
According to the concept of self-expression and its drivers, if a software application can help people to 1) express their specific identity -- \emph{Ideal Self}, and 2) distinguishing themselves from others (e.g. status among peers) -- \emph{Public Self}, it gives them sense of self-expression which influences people's preference positively and can enhance their emotional attachment. 

\paragraph{Affiliation}

According to the Theory of Human Personality \cite{champoux2016organizational} (Section \ref{sec:emattach}), people may use a product or service to feel a sense of belonging to social groups or affiliate with others. People may also use a software system as it provides them with a platform to make a sense of belongingness so that they can form their social identity. Accordingly, if a software system can help people to belong to a group or enhance that part of the self that needs to feel connected, it will satisfy the people's desires for affiliation \cite{meschtscherjakov2014mobile}. For instance in a system to help elderly people feel more cared for \cite{miller2015emotion}, the older person may experience a stronger positive emotional attachment with an application that makes them feel more connected to their family members, friends, carers, or other social groups.

\paragraph{Pleasure}
According to what Jordan \cite {jordan2002designing} argued within the context of product design, in addition to function and usability, people seek pleasure from a software system that they use. Therefore based on what we discussed in Section \ref{sec:emattach}, we can infer that if people experience pleasure in using a software system in any form of physical, social and ideological, they feel emotionally attached to the software system. In other words, if a software system provides sensory and enjoyable activity, it causes emotional attachment. 

\paragraph{Memories}
We can extend the Mugge's argument \cite{mugge2010product} regarding the role of memories in attaching consumers to a product emotionally (Section \ref{sec:emattach}), to the software engineering domain. Accordingly, if a software system can help remind people of the past, a specific occasion, an important person, create sense of personification, convey cultural-religious meanings or cause feelings of nostalgia, it may emotionally attache people to the software system. For example, reminders about family and friends' birthdays or bringing up old photos posted to a social networking size.

We call the four above-mentioned sources of emotional attachment in EAF as \emph{primary themes} and the sub-themes as \emph{secondary themes}. A brief description of each emotional attachment primary and secondary theme is found in Table \ref{definition}. It is clear that the primary themes are intertwined, rather than independent. For instance, in designing a mobile app for educating people to adopt low carbon products, gamification can stimulate the pleasure and self-expression of people to connect them to other environmentally friendly people or green communities (affiliation).  Based on the definitions presented in Section \ref{sec:emattach} and Table \ref{definition} we realize that `affiliation', `social pleasure' and `public self' are all closely connected. While their definitions and underlying theories are distinct as illustrated in the framework, it is expected that goals that are related to one have a high chance of also being related to the other two due to the fact that they are all related to social drivers of emotional attachment. The dashed oval in Figure \ref{fig:fig3} represents this proximity between these three themes. 


\begin{table}{}
\centering
\caption{Themes Definitions and Clues}
\label{definition}

\begin{tabular}{lp{0.6\textwidth}p{0.07\textwidth}}
\toprule
\textbf{Themes} &  \textbf{Definition}/\textbf{Clue} & \textbf{Source(s)} \\ \midrule
\textbf{Self-expression} & \textbf{Definition:} A person's aspiration to feel a distinct personal identity from others through expressing their personal identity such as feelings, thoughts or ideas  & \cite {mugge2008product,  ball1992role, kleine1995possession} \\
 &\textbf{Clue:} Any evidence regarding a person's aspiration to have a distinct personal identity from others & \\[1mm]

~~--- Ideal-self  & \textbf{Definition:} A person's aspiration to feel like a person they would imagine to be & \cite {sirgy1982self} \\
& \textbf{Clue:} Any evidence regarding a person's aspiration to have particular characteristics, skills, etc. & \\[1mm]

~~--- Public-self  &  \textbf{Definition:} A person's aspiration regarding the image they would like others to have of them & \cite {sirgy1982self} \\ 
& \textbf{Clue:}  Any evidence regarding a person's aspiration to create a particular image from themselves in the eyes of others & \\[1mm]

\midrule
\textbf{Affiliation}  & \textbf{Definition:} A person's aspiration to feel a sense of relationship with others, and a belonging to social groups & \cite{champoux2016organizational, kleine1995possession} \\
& \textbf{Clue:} Any evidence regarding a person's aspiration for connection, joining, associating, and involving with others & \\[1mm]
\midrule
\textbf{Pleasure}  & \textbf{Definition:} A person's aspiration to feel happiness satisfaction and enjoyment & \cite {Mugge, jordan2002designing, maclachlan2009exploring, maclachlan2009let} \\
& \textbf{Clue:} Any evidence regarding a person's aspiration of having pleasure and happiness as the primary or most aim of their life & \\[1mm]
~~--- Physical pleasure  & \textbf{Definition:} A person's aspiration to feel pleasure by means of a sensory perception (five senses) & \cite {jordan2002designing} \\ 
& \textbf{Clue:} Any evidence regarding a person's aspiration to feel sense of realism in using technology & \\[1mm]
~~--- Social pleasure & \textbf{Definition:} A person's aspiration to feel pleasure resulting from social interaction or social status/image & \cite {jordan2002designing} \\
& \textbf{Clue:} Any evidence regarding a person's aspiration to have or improve their social interaction and social status & \\[1mm]

~~--- Ideological pleasure  &  \textbf{Definition:} A person's aspiration to feel pleasure as a consequence of supporting their values and beliefs & \cite {jordan2002designing} \\ 

& \textbf{Clue:}  Any evidence regarding a person's aspiration to follow their personal values, beliefs and attitudes & \\[1mm]
\midrule
\textbf{Memories} & \textbf{Definition:} A person's aspiration to feel a sense of their past, happy moments,  people, occasions and places that are important for them & \cite {klein2012memory, csikszentmihalyi1981meaning} \\
& \textbf{Clue:} Any evidence regarding a person's aspiration for constructing and maintaining a sense of the past, happy moments in life or remind people, occasions, places etc. & \\[1mm]

\bottomrule
\end{tabular}
\end{table}

\begin{figure}[]
\small
\centering
\includegraphics [scale=0.6]{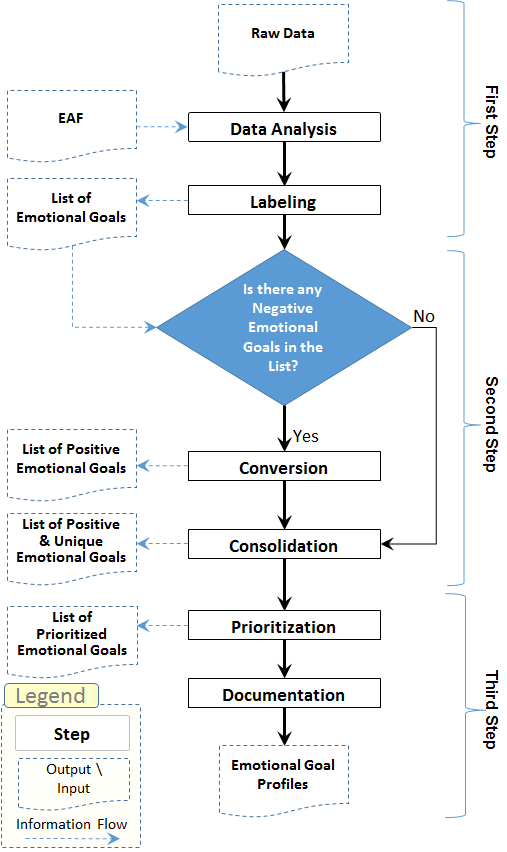}
\caption{Process Model} 
\label{fig:fig5}
\end{figure}

\subsection{Process Model}
\label{sec:method}



The main purposes of proposing the EAF is to develop a technique for helping system analysts understand the underlying drivers behind people's emotional statement and breaking them down to illustrate the key emotional goals of a software system. For this purpose in this section we outline a process model for understanding emotional goals by using the EAF. As shown in Figure \ref{fig:fig5}, the required input to the process model is data. Data collection can be achieved using any proper qualitative, quantitative or mixed data collection approach such as interviews, surveys, observations, focus groups, ethnography, documents and records \cite {neuman2005social}.

\subsubsection{First Step: Derive Emotional Goals}
\label{sec:firststage}

\medskip 

\begin{center}
\begin{tabular}{@{}lllll@{}}
\toprule
\textbf{Task} & \textbf{Input} & \textbf{Technique} & \textbf{Output} & \textbf{\begin{tabular}[c]{@{}l@{}}Terminating Condition\end{tabular}} \\ \midrule
Data Analysis & \textit{\begin{tabular}[c]{@{}l@{}}Raw Data \\ \end{tabular}} & \textit{\begin{tabular}[c]{@{}l@{}}EAF\end{tabular}} & \textit{\begin{tabular}[c]{@{}l@{}} Emotional Goals \end{tabular}} & \textit{\begin{tabular}[c]{@{}l@{}} Emotional Goals Saturation\end{tabular}} 
\\
Labeling & \textit{\begin{tabular}[c]{@{}l@{}}Emotional Goals \end{tabular}} & \textit{\begin{tabular}[c]{@{}l@{}}$-$\end{tabular}} & \textit{\begin{tabular}[c]{@{}l@{}}List of Emotional Goals\end{tabular}} & \textit{\begin{tabular}[c]{@{}l@{}} No Emotional Goals without Label \end{tabular}} 

\\ \bottomrule
\end{tabular}
\end{center}

\medskip

The purpose of the first step is to analyse the data and extract the relevant information related to people's emotional goals.  Although system analysts may be able to analyse the data for exploring the emotional goals without using the EAF, we argue that the EAF provides a more systematic and complete way of performing this analysts, helping to provide insights to analysts and reducing the risk of  overlooking key emotional goals. 

In the first step, the \emph{emotional attachment themes} are used to analyse data. These are: \emph{Self-Expression}, \emph{Affiliation}, \emph{Pleasure} and \emph{Memories}. System analysts extract 
emotional statements, which are representative data points that help to inform the derivation of emotional goals. These can be from sources such as interview quotes, feedback data, etc. To extract these, we use the themes in the EAF. Such as step is non-trivial, however, in Table \ref{definition}, for each theme, we provide clues for helping to extract these statements.

After that, system analysts need to define core emotional goal(s) in each of the derived emotional statements and label them meaningfully. The labels should be concise and without ambiguity to give the reader a sense of what the associated emotional goal is about. 
For instance, `homeless people would like to have social interaction and relationship with others' is a sample of emotional statement. The core emotional goal in this emotional statement can be defined as `Connected'. 
This step will be continued until the emotional goals are fully explored and labeled and system analysts reach emotional goal saturation which means no more emotional goals are emerging from the emotional statements.

\subsubsection{Second Step: Emotional Goal Conversion and Consolidation}
\label{sec:secondstage}

\medskip 

\begin{center}
\begin{tabular}{@{}lllll@{}}
\toprule
\textbf{Task} & \textbf{Input} & \textbf{Technique} & \textbf{Output} & \textbf{\begin{tabular}[c]{@{}l@{}}Terminating Condition\end{tabular}} \\ 
\midrule
Conversion & \textit{\begin{tabular}[c]{@{}l@{}}List of Emotional \\ Goals\end{tabular}} & \textit{Paraphrasing} & \textit{\begin{tabular}[c]{@{}l@{}}List of Positive \\ Emotional Goals\end{tabular}} & \textit{\begin{tabular}[c]{@{}l@{}}No Negative  Emotional Goal \\in the List\end{tabular}} \\
\midrule
Consolidation & \textit{\begin{tabular}[c]{@{}l@{}}List of Positive \\ Emotional Goals\end{tabular}} & \textit{Combination} & \textit{\begin{tabular}[c]{@{}l@{}}List of Positive \&\\ non-repetitive Emotional Goals\end{tabular}} & \textit{\begin{tabular}[c]{@{}l@{}}No Repetitive Emotional Goal \\ in the List\end{tabular}} \\ 
\bottomrule
\end{tabular}
\end{center}

\medskip

As discussed in Section \ref{sec:background}, people express their emotional goals in  positive, negative, and neutral forms. However, we argue that negative and neutral emotions should be converted to positive emotions as a preparatory step for goal consolidation. It also is important because we believe it is difficult to design, implement, and track for a negative/neutral emotional goal, compared to a positive emotional goal. 
For instance `having a cloudy mind' as a negative emotional feeling may be converted to `sense of clarity' as a positive emotional goal. 

The step one is creative and we should avoid being critical, letting the creative juices flow to extract emotional goals. While the second step should be critical with the aim of merging goals are trying to eliminate ideas that are incorrect or not useful. As a result, the second step produces a list of non-repetitive emotional goals by asking whether there is any overlap between the extracted emotional goals. If the derived emotional goals have similarity or overlap then they will be combined and represented by a single encompassing emotional goal. 

As an example, \emph{"sense of relation"} and \emph{"sense of association"} are two emotional goals that have been achieved in the first step (\ref{sec:firststage}). However, both of these emotional goals refer to the same concept, \emph{"connection"}. Accordingly, we can merge these two emotional goals and replace them with a non-repetitive goal like \emph{"connected"}.
The output of this stage is a list of non-repetitive and positive emotional goals. 



\subsubsection{Third Step: Develop Emotional Goal Profiles}
\label{sec:thirdstage}

\medskip 


\begin{center}
\begin{tabular}{@{}lllll@{}}
\toprule
\textbf{Task} & \textbf{Input} & \textbf{Technique} & \textbf{Output} & \textbf{\begin{tabular}[c]{@{}l@{}}Terminating \\ Condition\end{tabular}} \\ \midrule 

Prioritization & \textit{\begin{tabular}[c]{@{}l@{}}List of Positive \& \\ non-repetitive Emotional Goals\end{tabular}} & 
\textit{\begin{tabular}[c]{@{}l@{}}Percentage of \\ Frequency (POF)\end{tabular}} & \textit{\begin{tabular}[c]{@{}l@{}}List of Prioritized \\ Emotional Goals\end{tabular}} & \textit{\begin{tabular}[c]{@{}l@{}}No non-prioritized   \\  Emotional Goal \\ in the List\end{tabular}} 
\\
\midrule
Documentation & \textit{\begin{tabular}[c]{@{}l@{}}List of Prioritized  \\ Emotional Goals\end{tabular}} & \textit{\begin{tabular}[c]{@{}l@{}}User Emotional \\ Story \end{tabular}} & \textit{\begin{tabular}[c]{@{}l@{}}Emotional Goal\\ Profile \end{tabular}} & \textit{\begin{tabular}[c]{@{}l@{}}No Emotional \\ Goal in the List\end{tabular}}
\\
\bottomrule
\end{tabular}
\end{center} 

\medskip

The first task in step three is prioritization. Like functional and quality goals, there are many reasons that some emotional goals may be selected over others for analysis and system development. For instance, software system cannot consider all users' goals in the first iteration due to budget and time limits. Whilst in an incremental system development strategy, system analysts consider the most important requirements at first and then extend the system's boundary based on the other requirements. There are different techniques for prioritizing users' goals such as MoSCoW, Paired Comparison and 100-Point  \cite{hatton2008choosing}. In this research it is our assumption that the frequency of emotional statement is a good representation of the importance of users' emotional goals.
As a result, a simple method for prioritizing each emotional goal is checking how often the related emotional statement has been reported by stakeholders. It is possible to elicit priorities from people as well, but we believe that the count is a reasonable indicator. For this purpose we calculate the Percentage of Frequency (POF) by using the following: 

	\begin{equation}\label{eq:1}
		POF = \frac{Frequency~of~each~Emotional~Statement}{MAX~Frequency}
	\end{equation} 

In this study we propose a prioritization scales as shown in Table \ref{PS}. Based on the POF, a higher priority indicates that the related emotional goals have higher importance and priority and system analysts needs to set a higher priority to those emotional goals in designing and developing the software application.

\begin{table}[ht]
\centering
\caption{Prioritization Scales}
\label{PS}
\begin{tabular}{ll}
\toprule
POF (\%)                           & Priority \\
 \hline
$POF  \leq 15\%$            & Low      \\
$15\% < POF < 75\%$ & Medium   \\
$POF \geq 75\%$          & High    \\
\bottomrule
\end{tabular}
\end{table}

\begin{table}[ht]
\centering
\caption{Sample Emotional Goal Profile}
\label{EGP}
\footnotesize
\resizebox{\textwidth}{!}{%
\begin{tabular}{ll}
\toprule
\multicolumn{2}{c}{\textbf{Emotional Goal Profile}} \\
\midrule
Emotional Statement: & Homeless people would like to have social interaction with and relate to others.\\
Emotional Goal: & Connected \\
Theme(s): & Ideal Self, Public Self, Affiliation, Social Pleasure \\
Priority: & High \\
Emotional User Story (EUS): & As a homeless person, I want to have social interaction and relationship \\ & with others so that I feel connected.\\

\bottomrule
\end{tabular}%
}
\end{table}

The second task in step three is documentation. 
The output of the process model should be documented in a format that can be used as an input to existing software engineering techniques in order to proceed the derivation process into a final requirements specification. The output of the documentation task is represented in a format entitled \emph{Emotional Goal Profiles}. An emotional goal profile contains a summary of an emotional statement, emotional goal, key emotional themes in EAF that are associated with this emotional goal, the emotional goal priority and user emotional story. 

As a user story is a frequently used artifact in software engineering domain for representing requirements \cite{paetsch2003requirements}, we adapt this format for our purposes and generate \emph{User Emotional Stories}. It is our assumption in this study that the stories facilitates sensemaking the abstract emotional goals in a way that software engineers are familiar with, i.e. user story. 
The general format of a user emotional story is based on traditional user stories with a specific the focus and composition including emotional statement and emotional goals. In this study we use the following format for representing the stories:

\begin{center}
\emph{As a $<$type of user$>$, I want $<$emotional statement$>$ so that I feel $<$emotional goal$>$.}
\end{center}


An example user emotional story can be found in Table \ref{EGP}. As Table \ref{EGP} shows, having an emotional user story like this is more informative for a system analyst than simply a statement like ``designing for Joy".



\section{Evaluation}
\label{sec:evaluation}

Any artifacts in information systems and software engineering are developed to improve task performance in two ways: (i) improving the quality of the result and, (ii) reducing effort required to complete the task \cite{moody2003method}. Accordingly, the aims of our evaluation in this research are as below:

\begin{itemize}
    \item Measuring the proposed framework's effectiveness: to what extent can the proposed framework help system analysts to achieve better perception regarding stakeholders' emotional goals?
    
    \item Measuring the proposed framework's efficiency: to what extent can the proposed method reduce effort required to achieve better perception regarding stakeholders' emotional goals?
\end{itemize}

For measuring the effectiveness of the proposed framework, three quality measures including \emph{completeness}, \emph{correctness} and \emph{consistency} (3Cs) were adopted by reviewing the literature \cite{pennotti2009evaluating, zowghi2002three, lee2014software, pressman2005software}. In this paper, we define these quality measures as below:


\begin{description}

\item [Completeness:] The proposed framework and process model lead to complete analysis, if the variety of required emotional goals have been specified.

\item [Correctness:] The proposed framework and process model lead to correct analysis, if it represents the accurate and correct reflection of required emotional goals.

\item [Consistency:] According to the definition of consistency \cite{trochim2001research}, the proposed framework and process model leads to consistent results if (i) the achieved results be consistent within itself (internal consistency) and, (ii) the same result can be repeatedly derived (external consistency) \cite{leung2015validity}. 

Although contextualizing people's emotional goals is subjective and different factors like context, system analyst's experience, background and expertise can affect the results, we believe that (i) the high level insight generated by the proposed framework regarding the required emotional goals should be consistent and, (ii) different people will be able to consistently derive required emotional goals. It is our assumption that if different people use the EAF for the same data set, regardless of the term used to describe the emotional goals, the general perception will be the same.

\end {description}

For measuring to what extent the proposed framework and its process model reduce effort required in understanding and capturing the required emotional goals,  different metrics such as time to complete a task, effort to learn and use the method \cite{moody2003method} can be used.    
In this research by reviewing the literature \cite{moody2003method, davis1989perceived, fitzgerald1991validating, mendoza2010learnability, mendoza2010software} following metrics were used for measuring the method's efficiency: 

\begin{itemize}
    \item \textbf{Time:} this metric measures the time taken to complete the task by using the proposed framework and process model.
    \item \textbf{Perceived Ease of Learning:} this metric measures to what extent the proposed framework and process model would be easy to learn. 
    \item \textbf{Perceived Ease of Use:} this metric measures to what extent the proposed framework and process model would be easy to use. 
    \item \textbf{Perceived Usefulness:} this metric measures to what extent the proposed framework and process model would be effective in achieving its intended objectives.
    \item \textbf{Intention to Use:} this metric measures to what extent a person intends to use the proposed framework and process model.
\end{itemize}

For measuring the effectiveness and efficiency criteria by people independent of the authors, we recruited four independent domain experts in our application domain. The case study is described in detail in Section \ref{sec:casestudy}.  
Two of the domain experts are exceedingly well versed in the domain of developing digital applications for supporting and assisting homeless people in Australia and were heavily involved in design and developing the initial version of our application. The other two domain expert were also part of the project and had extensive experience in designing and developing socio-technical software applications. 
The feedback from the experts that worked directly in design of this application is valuable given the wealth of experience these individuals possessed. Domain experts worked independently, except for times when they required clarifications for doing the tasks. 

For the semi-controlled experiment, $12$ participants were also recruited from a range of experience and expertise in software engineering domain and administered surveys to each. All the participants had master's degree or higher in the field of software engineering with three to five years working experience. Nine participants had specific training or experience in requirements engineering.
Table \ref{sumofeval} shows a summary of evaluation techniques used in this study. 

In the following sections, 
we discuss the case study and semi-controlled experience. For each activity, an overview of results will be discussed. The last part of this section is dedicated to reviewing some of the lessons that we learned. 

\begin{table}[ht]
\centering
\caption{The Summary of Evaluation Techniques}
\label{sumofeval}
\begin{tabular}{llllc}
\toprule

\textbf{Evaluation Goal} & \textbf{Evaluation Method} & \textbf{Evaluation Criteria}  & \textbf{Coverage} & \multicolumn{1}{l}{\textbf{Participant No.}} \\ \hline
\multirow{3}{*}{\textbf{Effectiveness}} & Case Study \& & \multirow{3}{*}{3Cs}  & Domain experts  & 4  \\ 
& Semi-controlled & & &  \\ 
& Experiment &  & Software engineers & 12  \\ \hline
\multirow{5}{*}{\textbf{Efficiency}} & & Time  &  \multirow{5}{*}{Software engineers} & \multirow{5}{*}{12} \\ 
& Semi- & Perceived Ease of Learning & &  \\ 
& controlled & Perceived Ease of Use & &  \\ 
& Experiment & Perceived Usefulness & &  \\ 
& & Intention to Use & &  \\ 
 \bottomrule
\end{tabular}%
\end{table}

\subsection{Case Study Analysis}
\label{sec:casestudy}
For doing the case study analysis, one of the authors of the present paper applied EAF in a real-world case study, developing a mobile application called \emph{Ask Izzy}. 

The pathways to homelessness in Australia are varied and this leads governments and industries to fund a range of services to support people who are homeless or at risk of homelessness.
Based on the situation of homeless people, a mobile phone can play a significant role in their life since in crisis, as even without credit, they can call emergency services or use free wi-fi. Survey statistics  show that in 2016, 95\% of homeless people in Australia have a mobile phone and majority of them (77\%) are equipped with smart phones. It is estimated that 42\% of the population of homeless or at risk people in Australia are young people (24 years and under) who are more likely to own a smart phone \cite {infoxchange}.

In Australia, there are more than 1200 government-funded specialist homelessness services and 300,000 health and welfare community support services \cite {infoxchange}. This abundance and diversity of services makes it difficult for those affected by homelessness to find the service they may need.
A system that can connect people to the help they need will be beneficial. For this reason, Infoxchange, a not-for-profit social enterprise in Australia, proposed a website application called \emph {Ask Izzy}\footnote {See \url{http://www.askizzy.org.au}} to connect homeless or people at risk  with essential support services and important information. This project is in collaboration with founding partners Google, REA Group and News Corp Australia.

Ask Izzy provided us with the opportunity of evaluating our proposed framework using interview data from people who had a genuine investment in the software application under development. The case study brought a rich representation of a variety of complex emotional goals drawn from lifetime experiences. Additionally, this case study provided the realistic setting of an industrial case study, along with the realistic pressures and constraints that come with an agile software development process. This rich socio-technical testbed would not have been possible to replicate in any artificial / toy study setup.

\subsubsection{Method}
\label{case method}

The initial version of \emph {Ask Izzy} was analysed using People-Oriented Software Engineering (POSE) model \cite{miller2015emotion} to understand homeless people's emotional goals in designing the initial version of \emph {Ask Izzy}. For data gathering they used semi-structured interviews with 22 participants including 15 homeless people and 7 social support workers. Each interview taking between 60-70 minutes, each participant was interviewed once. The interviews were focused on homeless people's requirements for designing a homelessness social support application entitled \emph{Ask Izzy}. 

In each interview some general questions were asked. These questions were based on the following themes: 1) what should the technology (web application) do for you?; 2) How should it be?; and 3) How do you want to feel? We did not ask these three questions directly, but these were just the themes around which we structured our questions.
For those participants with experience in using technology, we also asked questions regarding problems they experienced using technology. The interview data was analysed by using the POSE 
approach to extract and model the key emotional desires and goals by different stakeholders. All interviews were recorded and transcribed. 

In this study, we used the existing set of interview data at hand and used the following steps based on the proposed process model (Figure~\ref{fig:fig5}):

\begin{itemize}
    \item Based on the first step of this process model, we used the EAF to analyze the interview transcripts. The  interview  data  was  analyzed to specifically extract emotional goals. The analysis involved careful reading of the interview transcripts by using the themes  guideline (Table \ref{definition}) to achieve a better perception regarding the homeless people's emotional goals.
    \item Based on the second step of the proposed process model as explained in Section \ref{sec:secondstage}, we transformed negative emotional goals to positive emotional goals by paraphrasing them and then combined the similar ones. 
    \item Based on the third step of the proposed process model as discussed in Section \ref{sec:thirdstage}, for each  emotional  goal, POF was calculated and emotional goal profiles were prepared.
\end{itemize}



\paragraph{Emotional Goals}
One of the authors analyzed  each  interview transcripts by using the EAF themes' definitions. After that, the authors of the present paper defined the core emotional goals in each of the derived emotional statements and labeled them meaningfully. At the end of this step, $117$ labeled emotional goals were achieved. Table \ref{sample2} shows samples of emotional statements (quotes) and associated emotional goals\footnote{The complete list is available at https://goo.gl/vK86i1}.
As we discussed in Section \ref{sec:firststage}, this step was continued until the authors convinced that no more emotional goals are emerging from the emotional statements.


Since all of the emotional goals in this case study were expressed in negative form\footnote{The complete list is available at https://goo.gl/vK86i1}, in the second step they were converted to positive emotional goals. Based on the similarity between the emotional goals, the research team were grouped the $117$ labeled emotional goals into $20$ positive and non-repetitive emotional goals. Figure \ref{fig:conclusion} in Appendix~\ref{append1} shows the breakdown of these goals into their respecitve primary and second themes.
In the next step, all the $20$ emotional goals were prioritized and emotional goal profiles were developed\footnote {The complete list of profiles  is available at https://goo.gl/nrwEg8}.
Table \ref{emotionalconcerns} presents a summary of results in the \emph{Ask Izzy} case study.


\begin{table}[ht]
\centering
\caption{Samples of Emotional Statements - \emph{Ask Izzy} Case Study}
\label{sample2}
\scriptsize
\begin{tabular}{p{0.9\textwidth}
}
\toprule

``It makes you feel bad when they interrogate you and make out like you're trying to evade the system" 
\\[1mm] 
``They'd tell you one thing and then you'd see them again a week   later and they'd tell you something completely different" 
\\[1mm] 
``What about the people they need to look after here?" 
\\[1mm] 
``You're in need, you're in crisis, you're in everything and you haven't... there's nothing for you that you can help." 
\\[1mm] 
``The thing that used to annoy me is they say ``Go online, register online and you can access all this stuff''. 
\\[1mm] 
``You have to be lucky of getting the right person in the right call otherwise you could be like turning around." 
\\[1mm] 
``I think there should be more education around that for the young.  I stress now because my son, he's ten" 
\\[1mm] 
``There's just so many things that you learn through other people" 
\\[1mm] 
``When people are homeless, they don't consider being ill is an occupational hazard, and they don't go to a doctor until they're really sick, and a lot of times it's really too late." 
\\[1mm] 
``You don't even talk to the other people. They don't even talk to you." 
\\[1mm] 
``That's what I'm trying to do is just change the attitude towards it" 
\\[1mm] 
``There is nothing worse than support worker saying: Well look, I think it's time for us... you know our three month period is up now, I've used all 66 hours with you, you know it's time to stand on your own feet." 
\\[1mm] 
``The collapse actually happened?"  But you don't, because you're shameful, you're not worthy, you're not a winner, all those sort of things resonating through you, so there's that element ..." 
\\[1mm] 
``If you put all the people that live in this country together, we're all talented in something" 
\\[1mm]
 \bottomrule
\end{tabular}
\end{table}


The third column in Table \ref{emotionalconcerns} represents the frequency of each emotional goal in the interview data. For example, number $6$ in row $15$ means that six different emotional statements in the interview data were related to the emotional goal of `I Feel people don't trust me as a homeless person'. 
A higher frequency in this column indicates that the related emotional codes of selected emotional concern was more common between interviewees, has higher importance and priority for homeless people.
As the third column in this table shows, between different emotional goals, `Sense of transparency and trust in the in-formation', and `Connected', `Supported' and `Assisted' are more common homeless people's emotional goals with the frequency of $15$, $13$ and $12$ respectively. 



\begin{table}[ht]
\centering
\caption{A Summary of Results -The EAF Approach}
\label{emotionalconcerns}

\begin{tabular}{cp{0.40\textwidth}cccccccc}
\toprule

\multirow{10}{*}{\textbf{No.}} & 
\multirow{10}{*}{\textbf{Positive \& Unique Emotional Goals}} & 
\multirow{10}{*}{\textbf{\rotatebox[origin=c]{90}{Frequency}}}&
\multicolumn{7}{c}{\textbf{Associated Themes in EAF}} \\ \cline{4-10}& 

\multicolumn{1}{c}{} & \multicolumn{1}{c}{} &  
 
\textbf{\rotatebox[origin=c]{90}{Ideal Self}}& 
\textbf{\rotatebox[origin=c]{90}{Public Self}}& 
\textbf{\rotatebox[origin=c]{90}{Affiliation}}& 
\textbf{\rotatebox[origin=c]{90}{Social Pleasure}}&
\textbf{\rotatebox[origin=c]{90}{Ideological Pleasure}}&
\textbf{\rotatebox[origin=c]{90}{Physical Pleasure}}&
\textbf{\rotatebox[origin=c]{90}{Memories}}
\\
\cline{4-10}& 

\multicolumn{1}{c}{} & \multicolumn{1}{c}{} &  \multicolumn{7}{c} {\textbf{Frequency of Emotional Goals}}

\\ \midrule

1 
& Sense of transparency and trust in the information & 15 &  &  & 6 & 15 &  &  & \\

2 
& Connected & 13 & 13 & 13 & 12 & 13 &  &  & \\

3 
&  Supported and assisted & 12 &  &  &  & 12 & 1 &  & 
\\

4 
& Empathized with and  listened to & 11 &  &  & 11 & 11 &  &  & \\

5 
& Respected & 11 & 11 & 11 &  & 11 &  &  &  \\

6 
& Trusted & 6 & 6 & 6 &  & 6 &  &  &  \\ 

7 
& Self-confident & 5 & 5 &  &  & 5 &  &  &  \\

8 
& Sense of privacy & 5 & 1 & 1 &  & 5 & 4 &  &  \\

9 
& Motivated and hopeful & 5 & 5 & 5 &  &  &  &  &  \\ 

10 
& Calm & 5 & 5 &  &  & 3 &  &  &  \\

11 
& Sense of  self-worth & 5 & 3 & 5 &  & 1 & 5 &  &  \\

12 
& Useful & 4 & 4 & 4 & 4 & 4 &  &  &  \\

13 
& Safe and secure & 4 & 4 &  & 3 & 4 &  &  &  \\

14 
& Sense of trust in others & 3 & 3 &  &  & 3 &  &  &  \\ 

15 
& Sense of priority & 3 &  & 3 &  & 3 &  &  &  \\ 

16 
& Positive & 3 & 3 &  &  &  &  &  &  \\

17 
& Empowered & 3 & 3 &  &  & 2 &  &  &  \\

18 
& Independent  & 2 & 2 & 2 &  & 2 &  &  &  \\

19 
&  In control & 1 & 1 &  &  &  &  &  &  \\ 

20 
& Sense of fairness and justice & 1 &  &  &  & 1 &  &  &  \\

\midrule
 & Sum & \textbf{117} & \textbf{69} & \textbf{50} & \textbf{36} & \textbf{101} & \textbf{10} & \textbf{0} & \textbf{0} \\ \bottomrule

\end{tabular}

\end{table}

The numbers in the remaining columns represent how many times related emotional statement for each emotional theme have been seen in the interview data. For instance, row $10$ in Table~\ref{emotionalconcerns} shows that interviewees mentioned the concept of this emotional goal $5$ times in different statements. In all cases, the core emotional goal of mentioned emotional statements was related to `Public Self' and `Ideological Pleasure', three times to `Ideal Self' and just in one case to `Social Pleasure'. A higher value indicates that 1) related emotional goal can be addressed more effectively through this driver; and 2) for those prospective people who have expressed this emotional goal, this driver has potential to create emotional attachment between people and software system.


From Table \ref{emotionalconcerns}, we can understand some useful facts that can help us to contextualize the emotional goals in this case study. 
For instance, according to the distribution of emotional goals in Table \ref{emotionalconcerns} we can infer that homeless people feel loss of control in their life when they become homeless. In other words, one of the emotional drivers for the potential users of \emph{Ask Izzy} will be to help users to regain this feeling of control in their every day life. As another example, Table \ref{emotionalconcerns} shows that \emph{Social Pleasure} is the most important driver for forming emotional attachment. In other words, the majority of the people's emotional goals can be addressed by considering \emph{Social Pleasure} in designing \emph{Ask Izzy}. Row $19$ in Table \ref{emotionalconcerns} also shows the complex range of associated drivers for feeling \emph{Connected}. The complexity of this emotional goal may be an accurate reflection of the complexity of the problems that were expressed by participants. Almost all of the emotional statements elicited for this goal were found to be associated with \emph{Public Self}, \emph{Affiliation} and \emph{Social Pleasure}. This high degree of overlap found between all three of these drivers supports what have has been argued in Section \ref{sec:ETF} regarding the correlation and internal relationship of these three drivers in forming people emotional attachment. 

The EAF has four primary themes: 1) Self-expression; 2) Affiliation; 3) Pleasure; and 4) Memories, as well as several secondary themes. By consolidating the number of different emotional goals associated to each theme, we understand which theme could potentially create more emotional attachment between people and systems. Table \ref{sum2} shows the number of different emotional attachment drivers for each theme.

\begin{table}[ht]
\centering
\caption{Summary of emotional attachment themes for each theme - EAF approach }
\label{sum2}


\begin{tabular}{lrrrrrrrr} 

\toprule
\textbf{} & 
\textbf{\rotatebox[origin=c]{90}{Ideal Self}}& 
\textbf{\rotatebox[origin=c]{90}{Public Self}}& 
\textbf{\rotatebox[origin=c]{90}{Affiliation}}& 
\textbf{\rotatebox[origin=c]{90}{Social Pleasure}}&
\textbf{\rotatebox[origin=c]{90}{Ideological Pleasure}}&
\textbf{\rotatebox[origin=c]{90}{Physical Pleasure}}&
\textbf{\rotatebox[origin=c]{90}{Memories}}
& \textbf{Sum}\\
\midrule
\textbf{Self-expression} & 69 & 50 &  &  &  &  &  & \textbf{119} \\
\textbf{Affiliation} &  &  & 36 &  &  &  &  & \textbf{36} \\ 
\textbf{Memories}  &  &  &  &  &  &  & 0 & \textbf{0} \\
\textbf{Pleasure} &  &  &  & 101 & 10 & 0 &  & \textbf{111}\\

\bottomrule
\end{tabular}
\end{table}

As 
Table \ref{sum2} indicates, the majority of the homeless people emotional goals were related to \emph{self-expression} and \emph{pleasure} ($119$ and $111$ respectively). In other words, the root of majority of the homeless people emotional goals goes back to people's needs for expressing their personal identity as well as seeking pleasure. The reason of importance for these two categories is not so complicated to understand if we consider the nature of the homelessness phenomenon. As a consequence of becoming homeless, homeless people become so dependent on others and are busy for the activities that they have to do to survive. As a result, they do not have this opportunity to do some activities to feel some degree of pride, self-respect, and self-worth, follow their dreams and interests or have some degree of entertainment. Therefore it is not a surprise if their main emotional goals are related to \emph{Self-expression} and \emph{Pleasure}.

\paragraph{Baseline}
As noted at the start of Section \ref{evalde}, some of the authors of the present paper applied the POSE method to understand and model  emotional goals for homeless people. The resulting experiences were included into the design of the initial version of \emph{Ask Izzy}, which is now a functioning application that has many real users\footnote{See \url{https://askizzy.org.au/}}. As part of the analysis, we use this version as a baseline against which to compare our method.
Table \ref{askizzypose} shows the summary of the proposed emotional goals defined using POSE approach. Reviewing the prior study results shows that the system analysts of the initial version of the \emph{Ask Izzy} found $44$ emotional statement and they suggested $12$ emotional goals for designing the \emph{Ask Izzy} \footnote{The complete list of emotional statements and results are available at https://goo.gl/oQBdn2}. 



\begin{table}[ht]
\centering
\caption{A Summary of Results -The POSE Approach}
\label{askizzypose}

\begin{tabular}{cp{0.35\textwidth}c}
\toprule
\textbf{No.} & 
\textbf{Positive \& Unique Emotional Goal} & 
\textbf{Frequency} \\ \hline



1 
& Good feeling / Good mode & 1   \\ 
2 
&  Sense  of  motivation and hopefulness & 1 \\ 
3 
& Priority  & 1 \\ 
4 
& Empathy / Sympathy & 2 \\ 
5 
& Self-confidence & 2 \\ 
6 
& Respected & 3 \\ 
7 
& Sense of calm & 3  \\ 
8 
& Safe  of  self-respect and self-worth & 3 \\ 
9 
& Sense  of  connection and Interaction & 5 \\ 
10 
& Sense of privacy & 5 \\ 
11 
& Sense  of  support and assistance & 5 \\ 
12 
& Sense  of receiving clear,  simple  and  accurate information & 13 \\ 
\midrule
& Total & \textbf{44} 
\\ \bottomrule

\end{tabular}

\end{table}

\subsection{Effectiveness Analysis}
\label{evalde}

For measuring the proposed framework effectiveness criteria (Table \ref{sumofeval}), we presented our emotional goals to four domain experts and gathered both qualitative and quantitative feedback. The following steps where conducted in the process of measuring the proposed framework's effectiveness based on the case study:

\begin{enumerate}

\item Two lists of homeless people's emotional goals including (i) what was derived for designing the initial version of \emph{Ask Izzy} by POSE (baseline) and, (ii) what the authors of the present paper derived by using the EAF (EAF version) were presented to the domain experts. 

\item The domain experts were asked to consider the two lists of emotional goals and answer the following questions\footnote{The questionnaire is available at https://tinyurl.com/ybm4th9v
}; (a) whether there were any additional emotional goals are needed in each list; (b) whether there were any incorrect emotional goals are in each list; and (c) whether there are any inconsistency within emotional goals in each list? They were also asked to reflect their attitudes towards the elicited emotional goals with other group members and determine which list better reflects the homeless people's emotional goals.

\item The domain experts were then asked to review the two lists and identify any emotional goals that were non-repetitive in one list compared to the other; that is, they could not map a goal in one list to a goal in another. Given step 2, this indicated a missing goal in the second list.




\end{enumerate}

\subsubsection {Results}

As we discussed at the start of Section \ref{sec:evaluation}, we investigated three evaluation criteria for measuring the effectiveness of the proposed framework; completeness, correctness and consistency. In the following, a summary of results for  measuring the effectiveness analysis will be discussed.

\paragraph{Completeness}
All the domain experts mapped all $12$ emotional goals in the baseline to an equivalent in some of the emotional goals that we achieved by using the EAF. 



As Table \ref{askizzypose} indicates, while, from the domain experts point of view, all the emotional goals in the baseline were achieved in the EAF version, we could achieve $8$ further emotional goals by using EAF in this study. The additional emotional goals cover something important that the previous did not. `Sense of fairness and justice', `Independent', `Useful' and `Trusted' are some samples of emotional goals that were uncovered by only this proposed method, but not in the POSE method. 



The domain experts answers to the open-ended questions provided some valuable points. One domain expert stated:
\begin{itquote}
It is useful to elicit [emotional] goals. ... using the framework is much more comprehensive.
\end{itquote}
\begin{itquote}
Although you cannot guarantee the completeness, it is a very good starting point for capturing [emotional goals].
\end{itquote}
Although domain experts acknowledged that the proposed emotional goals in EAF version appeared to be complete, they commented that for users who identified as Australian Aboriginal and Torres Strait Islanders, a `Sense of Belonging' was not adequately represented or prioritised in the final goal set. While this goal overlaps with existing goals, such as the goal of feeling `Connected' it was found to be distinct in terms of the underlying drivers. In this case, the `Sense of Belonging' was driven more by a need for cultural identification as opposed to a requirement for social support. 
The analysis of domain experts responses 
shows that the proposed framework has the potential to help system analysts to achieve a variety  of  emotional goals that users may have.

It is important to note that people's emotions change over time and it is therefore difficult to argue that the list of emotional goals are complete. However, these initial results appear promising given that the domain experts claim that the proposed framework has this potential to lead to achieve a good insight regarding the variety of emotional goals that users may have.

\paragraph{Correctness} 
The domain experts were asked to raise anything that they would like to add or delete in the list of emotional goals derived by using the EAF. Results show that both domain experts believe the homeless people's emotional goals have been elicited correctly by using the EAF. None of the domain experts raised a case that from their opinions is incorrect and needs to be changed or delete. It shows that from the domain experts' point of view the proposed method could lead to correct results. 

For instance, one domain expert stated: 

\begin{itquote}
All this [the list of emotional goals by EAF] is true certainly. They [the captured emotional goals] are all true and valuable and it is very hard to find any counterexample to argue.
\end{itquote}

Although domain experts had some suggestions for better wording of the emotional goals, by answering the related questions, domain experts approved the correctness of the elicited emotional goals.
The domain experts had some comments to the open-ended questions which provided some valuable points. From these results, we see several expected results. First, EAF can help system analysts and requirements engineers, who may not have strong domain expertise, to achieve a better insight about the people's emotional goals:

\begin{itquote}
It is useful to identify emotional goals particularly if someone has no experience. Even if you have experience, when you are designing an interview, it is useful.
\end{itquote}

\noindent 
Another domain expert also stated:

\begin{itquote}
The framework completely makes sense to me, ..., I think it is really helpful for analysts and designers to put together the questions that they might want to ask.
\end{itquote}

Understanding the people's emotional goals is so important for system analysts, and our results show that the EAF can help system analysts in this process to some extent. One domain expert mentioned:

\begin{itquote}
Having the categories helps you focus on the relationship between the emotional goals  to consolidate or remove ... for consolidating [the] emotional requirements, it is good to look within the proposed categories in the framework.
\end{itquote}

\noindent and

\begin{itquote}
Having this [framework] helps me to think in different ways. It could give you a hint and is helpful to create a pool of solution for each [emotional] category. It is good to have them in the background when you designing.
\end{itquote}

The results of the above mentioned analyses also support our hypothesis that the EAF can help system analysts to get a deeper insight regarding people's emotional goals. We attribute these results to proposed themes for analyzing the data, hierarchical layout of the emotional goals categories in EAF and, well-structured and easy to use process model. 

\paragraph{Consistency}

For measuring to what extent the achieved results are consistent within itself (internal consistency), we asked the domain experts to determine whether any inconsistency exists within the proposed emotional goals. There is a potential for inconsistency on the outputs related to (i) what is considered to be an emotional goal (ii) what theme each emotional goal is related to, and (iii) how these emotional goals are grouped.
None of the domain experts mentioned any inconsistency within the achieved emotional goals. It is an acknowledgment that the proposed framework and its process model did not lead into inconsistent emotional goals.  

\subsection{Semi-controlled Experiment}
The goals of this semi-controlled experiment are two fold. First, whether our proposed framework and its process model lead to complete, correct and consistent results when it is used by people independent of the authors (effectiveness). It is our hypothesis that the EAF is an effective technique for understanding and capturing emotional goals compared to other techniques. Second, whether our proposed method leads to performance improvement in understanding and capturing the emotional goals (efficiency). It is our hypothesis that the EAF is more time efficient, easier to use and easier to learn technique for understanding and capturing emotional goals compared to other techniques.

\subsubsection{Method}

To conduct the semi-controlled experiment, $12$ participants were recruited with a variety of experience and expertise in software engineering. Based on their responses in a screening survey, all of the participants had a master's degree or higher in the field of software engineering with three to five years working experience. Seven participants had specific  training in requirements engineering.

The participants were asked applied both (i) EAF and (ii) a rival method (i.e: POSE) to extract emotional goals from a subset of data of the \emph{Ask Izzy} case study, and measured the effectiveness and efficiency criteria (Table \ref{sumofeval}). 
The experiment used a within subject design, comparing two activities. In each activity, each participant was presented with the same sample of $12$ dialogue of what the homeless people in the \emph{Ask Izzy} case study have been mentioned during the interview sessions regarding their requirements and needs.
To mitigate the potential bias, the sample dialogues in each activity were selected randomly. The participants were asked to review the sample dialogues to get a deeper insight regarding homeless people's emotional goals. 
In the first activity, we introduced the participants the POSE to derive emotional goals from the 12 sample dialogues. 
In the second activity, the EAF was introduced to the participant, who was then asked to apply EAF to derive emotional goals and label them meaningfully.
At the end of each activity, we asked the participants in both groups to categorize the recommended emotional goals based on their similarity and form a list of non-repetitive and positive emotional goals. 
To avoid an order effect, we counterbalanced by asking half of the participants to use EAF in the first activity and half to use the POSE in the second activity; and then switched. Participants in worked independently, and requested to only ask questions to clarify process.


\subsubsection{Efficiency and Effectiveness Analysis}

For measuring the efficiency and effectiveness criteria (Table \ref{sumofeval}) in semi-controlled experiment, we measured different metrics between the EAF and POSE. For this purpose, we gathered both qualitative feedback and quantitative ratings. As such, the following steps where conducted:

\begin{itemize}

\item at the end of the each round, we timed how how long each participant spent to do his or her tasks in each round. There were no time limits on tasks and the average time spent by the participants was calculated in minutes.

\item for measuring the perceived ease of learning, perceived ease of use and perceived usefulness we administered a post-task survey at the end of the each round and asked participants in each group some open-ended qualitative questions about the used methods such as:(i) how comfortable they felt learning the method (easy to learn); (ii) how comfortable they would feel using the method for analyzing  emotional goals (easy to use); (iii) to what extent they found the used method useful (usefulness); and (iv) which method is their preference if they had to analyze people's emotional goals in the future? (intention to use)\footnote{The questionnaire  is available at https://tinyurl.com/y8xccehk}. 
In the designed questionnaire some general questions were also asked about participants general feeling about the used frameworks.

\end{itemize}

As it is difficult to objectively measure to what extent our proposed framework could help participants in semi-controlled experiment to achieve a better perception regarding people's emotional goals, we went back to the domain experts to assess the participants' output. 
We supplied the four domain experts with the sample dialogues used by the participants in semi-controlled experiment and two lists of emotional goals proposed by each participant in semi-controlled experiment corresponding to the two rounds. Domain experts did not know which emotional goals were analyzed using EAF and which were obtained using POSE. The domain experts were asked to do the same tasks as in the case study in Section~\ref{evalde}: identify incorrect goals, inconsistencies, and missing goals (incompleteness).

\subsubsection{Efficiency Results}

\begin{figure}[]
\small
\centering
\includegraphics [scale=0.5]{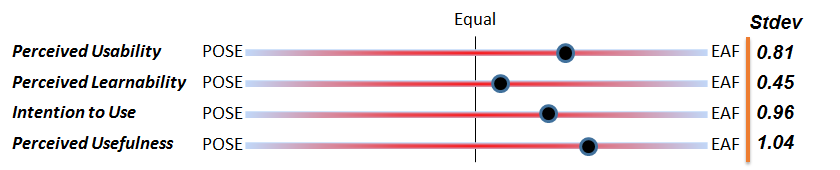}
\caption{The Summary of Efficiency Analysis Results} 
\label{fig:resultseff}
\end{figure}

Table \ref{semiexpresults} and Figure \ref{fig:resultseff} summarize the results for two experimental groups and evaluation metrics respectively.

\begin{table}[ht]
\caption{Summary of Semi-controlled Experiment Results }
\label{semiexpresults}
\centering
\begin{tabular}{p{1cm}cccc}
\toprule 

&\multicolumn{2}{c}{\textbf{Baseline}} &\multicolumn{2}{c}{\textbf{EAF}}  \\
\cmidrule(lr){2-3} 
\cmidrule(lr){4-5}

\textbf{Parti-cipant} & \textbf{\# of relevant} & \textbf{Time Spent} & \textbf{\# of relevant} & \textbf{Time Spent} \\
& \textbf{EGs*} & \textbf{(mins)} & \textbf{EGs} & \textbf{(mins)} \\

\cmidrule(r){1-5} 

P1 &5  &15.7  &7  &10.8   \\
P2 &5  &16.1  &8  &11.5   \\
P3 &6  &17.0  &9  &12.6   \\
P4 &5  &15.9  &7  &14.1   \\
P5 &4  &14.5  &8  &12.2   \\
P6 &3  &15.6  &7  &15.0   \\
P7 &5  &13.0  &8  &\phantom{1}9.8   \\
P8 &5  &14.9  &8  &13.2   \\
P9 &5  &18.1  &6  &15.2   \\
P10 &5  &13.7  &7  &10.4   \\
P11 &4  &16.0  &7  &11.8   \\
P12 &6  &18.0  &8  &11.3   \\
\cmidrule(r){2-3} 
\cmidrule(lr){4-5}

\textbf{Ave.} & \textbf{4.8} & \textbf{15.7}  & \textbf{7.5}  & \textbf{12.3} \\ 
\bottomrule
\multicolumn{3}{l}{p-value EG $= 0.00222$} &
\multicolumn{2}{l}{p-value for time $= 0.00222$}\\

 \multicolumn{2}{l}{*EG: Emotional Goal} 
\end{tabular}
\end{table}

One can see that on average, participants suggested more emotional goals in a faster average time by using the EAF . As Table \ref{semiexpresults} shows, the average of non-repetitive emotional goals suggested by the participants by using the POSE was $4.8$ while this figure was $7.5$ when the participants used the EAF. 
Comparing the time spent by the participants for analyzing the sample dialogues shows that the participants spent more time in order to elicit the emotional goal when they used the POSE ($15.7$ min) while these participants spent three minutes less on average for doing the same job when they used the EAF. 

For measuring to what extent the difference between the number of achieved emotional goals by using the POSE vs EAF and spent time are important, we conducted Wilcoxon signed-rank test \cite{sheskin2003handbook}. 
The Wilcoxon signed-rank test is a non-parametric statistical hypothesis test used to evaluate the difference between two treatments or conditions for the correlated samples. 
Accordingly, our null hypothesis is that there is no statistically significant differences between the number of emotional goals achieved by POSE and EAF. We conducted the Wilcoxon signed-rank test for a $0.05$ level of significance. Accordingly the null hypothesis will be rejected if the p-value  $\leq 0.05$.  
Table \ref{semiexpresults} shows the p-values for both number of emotional goals and time spent are $0.00222$. 
Therefore, the results are significant at the $5\%$ level.


Figure \ref{fig:resultseff} shows the mean and standard deviation of the responses for the questions associated with qualitative metrics. 
As Figure \ref{fig:resultseff} shows participants have a stronger preference for using
EAF, largely because it has a well-structured and easy to follow process model which lead to a more natural way to get the deeper insight regarding the emotional goals. 
Given that these measures are our proxies for usefulness, they imply that the EAF is easier to understand and use by participants.

The results of the above mentioned analyses also support our hypothesis that the EAF can help system analysts to achieve a better perception regarding people's emotional goals as the overall number of
suggested emotional goals per participant was higher and the times were also lower for the EAF.

The participants' responses to the quantitative questions provide further evidence of the EAF efficiency. Participants' responses support this argument that the EAF is more useful method than rival technique for achieving a better perception regarding people's emotional goals  in system analysis process. As Figure \ref{fig:resultseff} shows, all participants believe that the EAF is easy to learn and use for getting a deeper insight regarding people's emotional goals. For example, the following quotes are from two different participants:

\begin{itquote}
It [EAF] was straightforward for me. I could get it very soon. Maybe because it has been designed in more natural way.
\end{itquote}
    
\begin{itquote}
First I [was] confused by different terms but the hierarchy directed me clearly to learn the process. ... this step-wise process is helpful to repeat it again.
\end{itquote}

Figure \ref{fig:resultseff} shows clearly that is preference for the EAF. Participants emphasized their preference in their comments as well:

\begin{itquote}
\ldots I go for this [EAF], at least I have some themes and know what I am looking for in the dialogues.
\end{itquote}

\begin{itquote}
It [EAF] is absolutely my preference. I was like a blind person in the first round. Could find only $3$ [emotional goal] as I could not distinguish them form functional [goals] but the themes in the next round is doubled. It gave me more insight to extract even implicit emotional needs.
\end{itquote}

\begin{itquote}
I found it [EAF] useful to use it again as it led to consistent understanding. The categorization makes it easier to identify the emotional goals as it encapsulates the most common categories for people emotions.
\end{itquote}

However, one of the participants commented that he/she did not find the level of granularity useful for analyzing the emotional goals:

\begin{itquote}
They are too fine-grained. Sometimes it is hard to link an emotion to just one of those. Often a high-level themes would be better.
\end{itquote}

This participant also commented: 

\begin{itquote}
Might be applied and useful for analyzing the ordinary people's emotional goals, however, people with physical or mental illness or disability may expressed things differently.
\end{itquote}

Participants' comments show that they found the EAF useful for different reasons. Participants expressed it in different words:

\begin{itquote}
When you see the roots of the emotions and know their definitions you will have an organized mind which drives your attention and keep your focus on the concepts. \ldots Although I've never been homeless I think it [EAF] helps me to feel what a homeless feels.
\end{itquote}

\begin{itquote}
It makes your arguments regarding the validity of users' requirements more tangible and straightforward. It is nice to see the relation between the emotional goals and their psychological background.
\end{itquote}

\begin{itquote}
It provides a way to analyze emotions systematically and make the results easier to comprehend.
\end{itquote}

\begin{itquote}
It helped me to think of what could be an emotional goal which hard to distinguish sometimes using personal experience.
\end{itquote}



\subsubsection{Effectiveness Results}

The results of domain experts analysis show that our proposed method in this study could help participants  in semi-controlled experiment effectively. Table \ref{domainexperteval} shows the results of the domain experts analysis.

\begin{table} [ht]
\centering
\caption{Summary of domain experts evaluation results}
\label{domainexperteval}

\begin{tabular}{p{1cm}cccc}
\toprule

{Parti-cipant} & \multicolumn{2}{c}{\begin{tabular}[c]{@{}c@{}} Number of Unique \&\\Relevant EG*\end{tabular}} & \multicolumn{2}{c}{\begin{tabular}[c]{@{}c@{}} Number of  \\Inconsistent EG*\end{tabular}}\\
\cline{2-3}
\cline{4-5}
 & EAF & POSE
  & EAF & POSE
 
\\
\hline

P1 & 1.25 & 0.25 & 0.5 & 1.25\\
P2 & 2.00 & 0.00 & 0.25 & 0.75\\
P3 & 2.75 & 0.00 & 0.75 & 1\\
P4 & 1.50 & 0.50 & 0 & 1\\
P5 & 2.25 & 0.75 & 0.25 & 1.75\\
P6 & 2.75 & 0.75 & 0.25 & 0.5\\
P7 & 2.00 & 0.00 & 0.5 & 1.25\\
P8 & 1.25 & 0.50 & 0.5 & 1.25\\
P9 & 0.25 & 0.00 & 0.25 & 1\\
P10 & 1.50 & 0.50 & 0 & 0.75\\
P11 & 2.00 & 0.00 & 0.5 & 1.25\\
P12 & 1.50 & 0.25 & 0.25 & 1.25\\
\midrule
Ave. & 1.75 & 0.29 & 0.33 & 1.08\\
\bottomrule
\multicolumn{5}{l}{p-value for number of EG = 0.00222}\\ 
\multicolumn{5}{l}{p-value for number of inconsistent EG = 0.00222 }\\ 
\multicolumn{3}{l}{*EG: Emotional Goal}
\end{tabular}
\end{table}

\paragraph{Completeness and Correctness}
As Table \ref{domainexperteval} shows, from the domain experts point of view, participants in semi-controlled experiment could find more non-repetitive and relevant emotional goals goals by using the EAF compared with the rival technique.  
Although we acknowledge that achieving more non-repetitive and relevant emotional goals is not a benchmark for measuring the proposed method completeness and correctness, the uniqueness and relevance of achieved emotional goals support the claim that the EAF has succeeded 
to provide a deeper insight regarding the people's emotional goals. It is important to note, however, that the incompleteness comes directly from the domain experts view: they regarded the additional goals as valuable and not superfluous to the analysis.

For measuring to what extent the difference between the domain experts' opinions regarding the number of non-repetitive and relevant emotional goals proposed by the participants by using the EAF and POSE is important, we again conducted Wilcoxon signed-rank test.
Accordingly, our null hypothesis was that from the domain experts' perspective, there is no statistically significant differences between the number of non-repetitive and relevant emotional goals achieved by the participants by using the POSE and EAF. 

We conducted the Wilcoxon signed-rank test for a 5\% level of significance. Table \ref{domainexperteval} shows the p-value between the EAF and the baseline method for emotional goals is  $0.00222$, so we reject the null hypothesis, concluding that the results are significant at $5\%$.

\paragraph{Consistency}

Although all the participants used the same data set and the main findings in the semi-controlled experiment were adapted from well-defined framework (i.e.\ EAF), there is a potential for inconsistency within the emotional goals suggested by each participant (internal consistency) and between the emotional goals suggested by the participants (external consistency). 

As Table \ref{domainexperteval} shows, from the domain experts point of view, participants in semi-controlled experiment could find emotional goals with less inconsistency by using the EAF compared with the POSE. The results show that the average number of inconsistent emotional goals was more than tripled in the baseline method.

We again conducted Wilcoxon signed-rank test for a 5\% level of significance.
Table \ref{domainexperteval} shows the p-value between the EAF and the baseline method for emotional goals is 0.00222. It means that from the domain experts' point of view, the EAF could cause statistically significant improvement in proposing less inconsistent emotional goals by each participant comparing with the POSE.

For measuring to what extent different participants could able to repeatedly derive the same emotional goals (external consistency), different techniques such as Inter-Rater Reliability, Test-Retest Reliability, Parallel-Forms Reliability and Internal Consistency Reliability are available \cite{trochim2001research}. Among the mentioned techniques inter-rater reliability is a popular technique in the system development domain \cite {boudreau2001validation}. We used Cohen's Kappa values for measuring the inter-rater reliability to understand to what extent the proposed method in this research leads to consistent results. Cohen's Kappa statistical measurements range from $-1.0$ to $1.0$; larger numbers represent better reliability and smaller numbers near zero suggest agreement has happened by chance \cite{saeed2013computer}.
As we discussed earlier, the output of the EAF is subjective and different factors like context, system analyst's experience, expertise and background used for analysis can affect the results. However, it is our assumption that if different people use the EAF and its process model for the same data set, regardless of the term used to describe the emotional goals, the main idea will be the same.

As part of the process, one of the authors of the present paper analyzed the participants' responses and categorized the emotional goals based on their similarities. Then the Cohen's Kappa values for measuring the proposed method consistency was calculated.
Table \ref{kappa} shows the statistical values for the Cohen's Kappa index. As all the Cohen's Kappa values for proposed emotional goals are above 70\%, which is considered the minimum value for inter-rater agreement to be considered consistent \cite {boudreau2001validation}, we can say that the proposed method could help the participants to derive the same emotional goals. It supports our hypothesis that the proposed framework can lead to consistent results. 


\begin{table}[ht]
\centering
\caption{Consistency Analysis - Cohen's Kappa Values }
\label{kappa}

\small

\resizebox{\textwidth}{!}{%
\begin{tabular}{clllllllllllll}
\toprule
\multicolumn{2}{l}{\multirow{2}{*}{}} & \multicolumn{12}{c}{\textbf{Emotioanl Goals}} \\ \cline{3-14} 
\multicolumn{2}{l}{} & P1 & P2 & P3 & P4 & P5 & P6 & P7 & P8 & P9 & P10 & P11 & P12 \\ \cline{3-14} 
\multirow{12}{*}{\textbf{\rotatebox[origin=c]{90}{Emotional Goals}}}  
 & P*1 &  & 75.10 & 70.38 & 78.55 & 78.40 & 75.38 & 78.07 & 79.55 & 74.59 & 75.69 & 71.72 & 72.69 \\
 & P2 &  &  & 75.55 & 76.61 & 75.03 & 70.27 & 74.07 & 77.17 & 79.89 & 78.46 & 78.16 & 77.68 \\
 & P3 &  &  &  & 79.02 & 79.04 & 75.90 & 71.20 & 70.58 & 79.47 & 72.41 & 78.79 & 73.65 \\
 & P4 &  &  &  &  & 79.76 & 70.08 & 79.02 & 73.79 & 70.15 & 72.51 & 72.76 & 75.06 \\
 & P5 &  &  &  &  &  & 74.50 & 79.83 & 73.64 & 76.74 & 77.65 & 70.18 & 76.00 \\
 & P6 &  &  &  &  &  &  & 77.45 & 71.84 & 79.04 & 72.93 & 70.47 & 71.50 \\
 & P7 &  &  &  &  &  &  &  & 79.22 & 78.73 & 77.64 & 73.10 & 78.31 \\
 & P8 &  &  &  &  &  &  &  &  & 76.89 & 74.53 & 77.92 & 76.94 \\
 & P9 &  &  &  &  &  &  &  &  &  & 74.28 & 75.95 & 71.87 \\
 & P10 &  &  &  &  &  &  &  &  &  &  & 72.23 & 71.60 \\
 & P11 &  &  &  &  &  &  &  &  &  &  &  & 73.73 \\
\bottomrule
 \multicolumn{14}{l}{P*: Participant}

\end{tabular}%
}

\end{table}


\subsection{Lessons Learned}
In this section, we discuss two main lessons learned in using our purposed framework in the \emph{Ask Izzy} case study. As we used different approaches for understanding people's emotional goals in our previous studies \cite{miller2015emotion, miller2012understanding, mendoza2013role}, we present this by contrasting our experience in using the EAF and other approaches that we used before. 

\paragraph{Questioning} Instead of exploring the emotional goals by using inductive analysis, using the EAF and proposed themes helped us in achieving a better perception regarding homeless people's emotional goals. 
    Although in using inductive analysis techniques like POSE we assume that if system analysts search hard enough, they will find emotional goals, it doesn't say enough about how this happens. These approaches also do not deal with the complexities of interpretation of the emotional goals.
    Using the proposed clues (Table \ref{definition}) in the process of analyzing the \emph{Ask Izzy} case study helped us to better interpret homeless people's dialogues while prevented us from thinking just about their explicit emotional goals. 
    
    Using the EAF structure in analyzing and validating the emotional goals in our case study caused great mental concentration. The hierarchy structure of EAF was also helped the research team to validate the achieved emotional goals in a systematic manner and find the overlap between the suggested emotional goals. For instance, by using the EAF we found that some of the quality goals that were derived by using the POSE in designing the initial version of the \emph{Ask Izzy} also came up as emotional goals - mainly context-specific emotional goals. For example, feel safe and secure was derived as a quality goal while by using the EAF it was derived as an emotional goal  since it is the subjective attributes of people and not the property of software. 

\paragraph{Labeling}
    One of the lessons that we learned was the importance of using a clear and unambiguous label for each emotional goal. We found it important as subjective, vague or ambiguous labels for emotional goals caused confusion for stakeholders in understanding the emotional goals. 
    Using labels that have multiple definitions can lead to confusion and disagreement when system analysts come to specifying and interpreting requirements. The likelihood result of confusion in interpreting the emotional goals is analyzing them in a wrong way and designing a software system with wrong features.
    We used different tactics in this study for reducing ill-definition and misinterpretation of emotional goals. First, we asked the project team members to review the EAF hierarchy several times and reflect their interpretations of the emotional goals. Based on the team members' interpretations we changed some emotional goals labels. Second, we used the emotional goal profile to define exactly what a certain emotional goal means in the EAF hierarchy and how it should be interpreted. 

\subsection{Limitations}

There are some limitations in this study. First, we consider only one case study. This is the trade-off that we made when deciding to ground this evaluation in a realistic industrial case study.  Accordingly generalising the usability and usefulness of the proposed framework is limited and further case studies would be a logical next step for this research. Additional studies would improve the validity of the proposed framework for achieving a better perception regarding people's emotional goals. Second, the case study is subject to bias, since our initial understanding from the case study and the homeless people's emotional goals would be used for data analysis and comparison the results. 
Participants may also have an interest in our framework. This proclivity can lead to inaccuracies in their responses.
Third, the analyzed data in the industry case study in this research focused only around the homeless people and social support workers goals. While data from others stakeholders such governmental sector organizations and business partners may be necessary for further validation. A threat to semi-controlled experimental analysis is that we used time and number of elicited emotional goals that may not be entirely accurate for measuring the efficiency of the proposed framework.

\section {Discussion}

In this study, the research question, \emph {How can system analysts achieve a better perception  regarding  stakeholders'  emotional  goals  in  the  process  of  requirements engineering?} is addressed. Although emotional goals and requirements have received some attention in software engineering, to the best of our knowledge, the proposed process model in this study is the first method that helps system analysts to systematically focus on people's emotional goals. 
As we experienced it in practice, a set of emotional themes is a good starting point to understand emotional goals. In doing the industry case study in this research, the EAF and the related themes gave us better insight into the domain and a better understanding the homeless people's emotional goals. We started the data analysis with a well-developed analysis framework that facilitated the process of understanding the homeless people's emotional goals.

This study suggests that although people may have different emotional goals, the root cause of their emotional goals can always be found behind what causes people to emotionally attach to (or detach from) a software system. However, existing approaches overlook indirect drivers of emotional attachment. There are many social situations that would be overlooked if they are categorized in other frameworks. For instance, categorizing the social isolation or the goal of feeling connected in other frameworks would end up with something quite abstract like ``joy", which  misses the social breakdown, and would potentially result in different requirements. Accordingly, understanding the people's emotional attachment themes help system analysts to understand better the people's emotional goals.
In our study, we captured $117$ emotional goals from 130,000 words transcripts that subsequently formed $20$ non-repetitive and positive emotional goals (Table \ref{emotionalconcerns}) under the three main emotional themes: 1) \emph{Self-expression}; 2) \emph{Affiliation}; and 3) \emph{Pleasure}.

In the \emph{AskIzzy} case study, participants spoke about the stigma associated with being homeless. The negative perceptions that society has on homeless people can obviously have a negative impact on the way they feel about themselves. According to the results, homeless people feel loss of control in their life when they become homeless. Therefore they want to regain control their everyday lives. This study  reveals that homeless people generally feel an inability to follow their interest that they had previously. The results indicate that empowering the potential users of \emph{AskIzzy} in following their interests would be not only important for them as a characteristics  that  they  would  like  to  have,  but  also  it  is important for them as part of their social interaction and social status.

The distribution of emotional goals associated with \emph{Affiliation} and \emph{Social Pleasure} themes indicates that the clarity and trust of information is an important goal for potential users and in designing an application would be need to be considered. The results of this study also reveal that there is a high degree of overlap between Public Self, Affiliation and Social Pleasure themes. This fact supports the correlation and internal relationship of these three themes in the EAF in forming emotional attachment. As the results show, the framework guides system analysts to extract subtle references related to the goal of being accepted and \emph{Feeling Respected} without the need to hear a direct verbal reference to how the situation is related to the way they feel.

Findings from this study show that using the EAF to elicit emotional goals within semi-structured conversations appears to have great potential. Without such support system analysts are at risk of overlooking important emotional goals. 
One potential reason for eliciting a greater number of emotional goals by using the EAF is that it allows the developer to extract quotes that are indirect (as well as direct) references to emotional goals. A better understanding of different aspects of people's emotional goals provides valuable insights for system analysts to design successful software application. 

Reviewing the results raise the question of whether goals that contain a high number of themes should be divided into smaller goals, allowing them to become easier to understand and design for. The optimal level of granularity for goals is still unknown, however there are certainly multiple different aspects of these goals that need to be addressed within design.
The EAF may not capture all emotional goals, and further, the quality of the output is only as good as the data that backs them. However, we believe that the list of emotional attachment themes for each emotional goal directs the system analysts' attention to the aspects that they need to consider in designing a software system to address the people's emotional goals. Furthermore, the proposed framework in this study (EAF) can be used before system analysts start the interviews to drive their dialogues and preparing the questionnaire.

Finally, evaluations of our proposed framework in \emph{Ask Izzy} case study provide support that from domain experts point of view the EAF can lead to improved systems analysts understanding of the people's emotional goals. It is our assumption that it potentially leads to better software system design. Furthermore, the process model in this study gives this ability to system analysts to prioritize emotional goals based on their frequency and focus on those that have a higher priority. A higher priority indicates that 1) related emotional goals can be addressed more effectively through the specific emotional themes, and 2) for those people who have this emotional goals, which themes have more potential to create emotional attachment between people and software system. 

\section{Conclusion and Future Work}

Over the past decade, there has been a paradigm shift from designing software systems for satisfying functional requirements towards the applications that are trying to enhance people's quality of life. We argue that due to this, system analysts need to engage with emotional relationships between software systems and people. 
Based on the unstructured nature of emotional goals, systematic frameworks such the EAF can increase rigor and improve analysis of the domain. It also can stimulate system analysts' sensitivity to consider the emotional aspects that they might not have considered otherwise.
In this paper, we introduced a novel framework and process model for understanding people's emotional goals, based on the sound theory of emotional attachment. This framework is simple by design, but conveys the important part of the information of the domain. 

The contributions of the EAF and its process model are first to propose an analysis method to deal with people's emotional goals for complementing existing requirements engineering techniques. In the process of analysing the people's emotional goals, the EAF draws attention to people's emotional attachment themes and provides concrete guidelines for eliciting emotional goals. However, its accuracy in terms of the volatile and changeable nature of  people's emotions is a matter of future work. Although the output of process model (EUS) provides input to the requirements engineering process, converting them into design implications requires further actions which we have presented in our another study \cite{paper1}.


The proposed framework (i) encourages system analysts to consider emotional themes before starting the data gathering such as interviews or workshops; and (ii) sensitizes system analysts' minds by providing a set of taxonomies and themes to discover emotional goals. Furthermore, the Emotional User Story,by using a known structure of user story, gives a clear guidance to system analysts to know how to contextualize people's emotional goals. 

We applied this framework to a software application that aids homeless people finding support services. Our transcribed interviews consisted of more that 130,000 words. We found that the emotional attachment framework provided a useful and structured way to analyse and collate emotional goals. The proposed framework in this study helped to find the right level of detail. We believe that our proposed framework is at an appropriate level for understanding the stakeholders' emotional goals. Although the quality of what we analysed in the case study is only as good as the data that backs it, we believe that these results support our assumptions that having a view of emotional attachment themes can facilitate the process  of understanding emotional goals.



To bridges the gap between emotional goals elicitation and software system design process, we have developed a systematic method for discovering possible design solutions entitled \emph {Emotional Goal Systematic Analysis Technique} (EG-SAT)\cite{paper1}. 
In future work, we aim to use both EAF and EG-SAT as complementary methods in other POS domains such as public health and education software systems for further investigation of the proposed frameworks usability and usefulness. We encourage other researchers to use the proposed framework in software system domain for developing the proposed framework and as a method for validation and repeatability.

\subsection*{Acknowledgements}
The authors would like to take this opportunity to express their gratitude to Alex Lopez-Lorca and Diana Brown
for their participation in our domain-expert studies, and valuable feedback on the model. 
This research is funded by the Australian Research Council Discovery Grant DP160104083 \emph{Catering for individuals' emotions in technology development}. The first author is funded by a University of Melbourne MIRS scholarship and a  top-up from the
CRC for Low-Carbon Living grant \emph{Increasing knowledge and motivating collaborative action on Low Carbon Living through team-based and game-based mobile learning}.

\bibliography{wileyNJD-AMA}%

\begin{thebibliography}{100}
\providecommand \doibase [0]{http://dx.doi.org/}%

\bibitem{calvo2014positive}
Calvo RA, Peters D. {\it Positive computing: Technology for wellbeing and human
  potential}.
\newblock MIT Press .
\newblock 2014.

\bibitem{guinan1986development}
Guinan P, Bostrom RP. Development of computer-based information systems: A
  communication framework. {\it ACM SIGMIS Database} 1986\string; 17(3)\string:
  3--16.

\bibitem{dieste2008understanding}
Dieste O, Juristo N, Shull F. Understanding the customer: what do we know about
  requirements elicitation?. {\it Software, IEEE} 2008\string; 25(2)\string:
  11--13.

\bibitem{gonzales2011eliciting}
Gonzales CK, Leroy G. Eliciting user requirements using Appreciative inquiry.
  {\it Empirical Software Engineering} 2011\string; 16(6)\string: 733--772.

\bibitem{colomo2010study}
Colomo-Palacios R, Hern{\'a}ndez-L{\'o}pez A, Garc{\'\i}a-Crespo {\'A},
  Soto-Acosta P. A study of emotions in requirements engineering. In:  {\it
  Organizational, Business, and Technological Aspects of the Knowledge
  Society}Springer; 2010\string: 1--7.

\bibitem{clancy1995standish}
Clancy T. The Standish Group Report.  1995.

\bibitem{OASIG1995}
OASIG, [cited on-line: accessed 19/12/2015]
  http://www.it-cortex.com/Stat-Failure-Rate.html the\%20OASIG\%20Study\%20).
  1995.

\bibitem{whittaker1999went}
Whittaker B. What went wrong? Unsuccessful information technology projects.
  {\it Information Management \& Computer Security} 1999\string; 7(1)\string:
  23--30.

\bibitem{tichy2008business}
Tichy L, Bascom T. The business end of IT project failure. {\it Mortgage
  Banking} 2008\string; 68(6)\string: 28.

\bibitem{van2001interactive}
Van~Harmelen M. {\it Interactive system design using {OO} \& {HCI} methods};
  Addison-Wesley .
\newblock 2001.

\bibitem{platt2007software}
Platt DS. {\it Why Software Sucks--and what You Can Do about it}.
\newblock Addison-Wesley Professional .
\newblock 2007.

\bibitem{Dix:2003:HI:1203012}
Dix A, Finlay JE, Abowd GD, Beale R. {\it Human-Computer Interaction (3rd
  Edition)}.
\newblock Prentice-Hall, Inc. .
\newblock 2003.

\bibitem{1663532}
Brooks J. No Silver Bullet Essence and Accidents of Software Engineering. {\it
  Computer} 1987\string; 20(4)\string: 10-19.

\bibitem{bentley2002putting}
Bentley T, Johnston L, Baggo vK. Putting some emotion into requirements
  engineering. In:  {\it Proceedings of the 7th Australian workshop on
  requirements engineering}; 2002\string: 227--244.

\bibitem{draper1999analysing}
Draper SW. Analysing fun as a candidate software requirement. {\it Personal
  Technologies} 1999\string; 3(3)\string: 117--122.

\bibitem{gogueny1994requirements}
Gogueny JA. Requirements engineering as the reconciliation of technical and
  social issues. {\it Requirements Engineering: Social and Technical Issues,
  edited with Marina Jirotka, Academic Press} 1994\string: 165--199.

\bibitem{hassenzahl2001engineering}
Hassenzahl M, Beu A, Burmester M. Engineering joy. {\it Ieee Software}
  2001\string; 18(1)\string: 70.

\bibitem{krumbholz2000implementing}
Krumbholz Ma, Galliers J, Coulianos N, Maiden N. Implementing enterprise
  resource planning packages in different corporate and national cultures. {\it
  Journal of Information Technology} 2000\string; 15(4)\string: 267--279.

\bibitem{miller2015emotion}
Miller T, Pedell S, Lopez-Lorca AA, Mendoza A, Sterling L, Keirnan A.
  Emotion-led modelling for people-oriented requirements engineering: the case
  study of emergency systems. {\it Journal of Systems and Software}
  2015\string; 105\string: 54--71.

\bibitem{proynova2011investigating}
Proynova R, Paech B, Koch SH, Wicht A, Wetter T. Investigating the influence of
  personal values on requirements for health care information systems. In:
  {\it Proceedings of the 3rd Workshop on Software Engineering in Health
  Care}ACM. ; 2011\string: 48--55.

\bibitem{sutcliffe2010analysing}
Sutcliffe A, Thew S. Analysing people problems in requirements engineering. In:
   {\it 2010 ACM/IEEE 32nd International Conference on Software Engineering}.
  2. IEEE. ; 2010\string: 469--470.

\bibitem{miller2012understanding}
Miller T, Pedell S, Sterling L, Vetere F, Howard S. Understanding socially
  oriented roles and goals through motivational modelling. {\it Journal of
  Systems and Software} 2012\string; 85(9)\string: 2160--2170.

\bibitem{callele2006emotional}
Callele D, Neufeld E, Schneider K. Emotional requirements in video games. In:
  {\it Requirements Engineering, 14th IEEE International Conference}IEEE. ;
  2006\string: 299--302.

\bibitem{mendoza2013role}
Mendoza A, Miller T, Pedell S, Sterling L, others . The role of users' emotions
  and associated quality goals on appropriation of systems: two case studies.
  In:  {\it 24th Australasian Conference on Information Systems}; 2013.

\bibitem{Goguen:1994:RER:177970.184582}
Goguen JA. Requirements Engineering. In: Academic Press Professional, Inc.;
  1994; San Diego, CA, USA\string: 165--199.

\bibitem{salzer1999atrs}
Salzer H. ATRs (Atomic Requirements) used throughout development lifecycle. In:
   {\it 12th International Software Quality Week}; 1999.

\bibitem{parrott2001emotions}
Parrott WG. {\it Emotions in social psychology: Essential readings}.
\newblock Psychology Press .
\newblock 2001.

\bibitem{plutchik2003emotions}
Plutchik R. {\it Emotions and life: Perspectives from psychology, biology, and
  evolution.}
\newblock American Psychological Association .
\newblock 2003.

\bibitem{ekman2007emotions}
Ekman P. {\it Emotions revealed: Recognizing faces and feelings to improve
  communication and emotional life}.
\newblock Macmillan .
\newblock 2007.

\bibitem{lama2008emotional}
Lama D, Ekman P. {\it Emotional awareness: Overcoming the obstacles to
  psychological balance and compassion}.
\newblock Macmillan .
\newblock 2008.

\bibitem{cambria2012hourglass}
Cambria E, Livingstone A, Hussain A. The hourglass of emotions. In:  {\it
  Cognitive behavioural systems}Springer; 2012\string: 144--157.

\bibitem{smith2009critiquing}
Smith H, Schneider A. Critiquing models of emotions. {\it Sociological Methods
  \& Research} 2009\string; 37(4)\string: 560--589.

\bibitem{arnold1970perennial}
Arnold MB. Perennial problems in the field of emotion. In:  {\it Feelings and
  emotions: The Loyola symposium}Academic Press New York. ; 1970\string:
  169--186.

\bibitem{Desmet2002}
Desmet P. {\it Designing Emotions}.
\newblock Pieter Desmet .
\newblock 2002.

\bibitem{cornelius1996science}
Cornelius RR. {\it The science of emotion: Research and tradition in the
  psychology of emotions.}
\newblock Prentice-Hall, Inc .
\newblock 1996.

\bibitem{thoits1989sociology}
Thoits PA. The sociology of emotions. {\it Annual review of sociology}
  1989\string: 317--342.

\bibitem{sep-emotions-17th18th}
Schmitter AM. 17th and 18th Century Theories of Emotions. In:  Zalta EN.
  \kern-2pt, ed. {\it The Stanford Encyclopedia of Philosophy}; 2014.

\bibitem{sweet1999frog}
Sweet F. {\it Frog: form follows emotion}.
\newblock Thames \& Hudson London .
\newblock 1999.

\bibitem{de1990rationality}
De~Sousa R. {\it The rationality of emotion}.
\newblock MIT Press .
\newblock 1990.

\bibitem{ortony1990cognitive}
Ortony A, Clore GL, Collins A. {\it The cognitive structure of emotions}.
\newblock Cambridge university press .
\newblock 1990.

\bibitem{norman2005emotional}
Norman DA. {\it Emotional design: Why we love (or hate) everyday things}.
\newblock Basic books .
\newblock 2005.

\bibitem{cowie2001emotion}
Cowie R, Douglas-Cowie E, Tsapatsoulis N, et al. Emotion recognition in
  human-computer interaction. {\it IEEE Signal processing magazine}
  2001\string; 18(1)\string: 32--80.

\bibitem{rulla2003depth}
Rulla LM. {\it Depth psychology and vocation: A psycho-social perspective}.
\newblock Gregorian Biblical BookShop .
\newblock 2003.

\bibitem{desmet2007framework}
Desmet P, Hekkert P. Framework of product experience. {\it International
  journal of design} 2007\string; 1(1).

\bibitem{scherer2001appraisal}
Scherer KR, Schorr A, Johnstone T. {\it Appraisal processes in emotion: Theory,
  methods, research}.
\newblock Oxford University Press .
\newblock 2001.

\bibitem{Frijda2010handbook}
Lewis M, Haviland-Jones JM, Barrett LF. Handbook of Emotions. In:  {\it
  Handbook of Emotions}Guilford Press; 2008\string: 68--88.

\bibitem{11015023620150701}
Monti~Fonseca LM, Min~Lun T, Vilela~Dias DM, et al. Emotional design and its
  contributions to digital educational technology in health and nursing:
  integrative review.. {\it Revista de Enfermagem Referencia} 2015\string;
  4(6)\string: 141 - 149.

\bibitem{norman2013design}
Norman DA. {\it The design of everyday things: Revised and expanded edition}.
\newblock Basic books .
\newblock 2013.

\bibitem{guo2014emotional}
Guo F, Liu WL, Liu FT, Wang H, Wang TB. Emotional design method of product
  presented in multi-dimensional variables based on Kansei Engineering. {\it
  Journal of Engineering Design} 2014\string; 25(4-6)\string: 194--212.

\bibitem{bretherton1992origins}
Bretherton I. The origins of attachment theory: {J}ohn {B}owlby and {M}ary
  {A}insworth.. {\it Developmental psychology} 1992\string; 28(5)\string: 759.

\bibitem{whan2010brand}
Whan~Park C, MacInnis DJ, Priester J, Eisingerich AB, Iacobucci D. Brand
  attachment and brand attitude strength: Conceptual and empirical
  differentiation of two critical brand equity drivers. {\it Journal of
  marketing} 2010\string; 74(6)\string: 1--17.

\bibitem{thomson2005ties}
Thomson M, MacInnis DJ, Park CW. The ties that bind: Measuring the strength of
  consumers' emotional attachments to brands. {\it Journal of consumer
  psychology} 2005\string; 15(1)\string: 77--91.

\bibitem{sable1995pets}
Sable P. Pets, attachment, and well-being across the life cycle. {\it Social
  work} 1995\string; 40(3)\string: 334--341.

\bibitem{mick1990self}
Mick DG, DeMoss M. Self-gifts: Phenomenological insights from four contexts.
  {\it Journal of Consumer Research} 1990\string; 17(3)\string: 322--332.

\bibitem{giuliani2003theory}
Giuliani MV. {\it Theory of attachment and place attachment}.
\newblock na .
\newblock 2003.

\bibitem{scannell2010defining}
Scannell L, Gifford R. Defining place attachment: A tripartite organizing
  framework. {\it Journal of environmental Psychology} 2010\string;
  30(1)\string: 1--10.

\bibitem{slater2001collecting}
Slater JS. Collecting brand loyalty: A comparative analysis of how Coca-Cola
  and Hallmark use collecting behavior to enhance brand loyalty. {\it
  NA-Advances in Consumer Research Volume 28} 2001.

\bibitem{mende2011attachment}
Mende M, Bolton RN. Why attachment security matters how customers' attachment
  styles influence their relationships with service firms and service
  employees. {\it Journal of Service Research} 2011\string; 14(3)\string:
  285--301.

\bibitem{ainsworth1969object}
Ainsworth MDS. Object relations, dependency, and attachment: A theoretical
  review of the infant-mother relationship. {\it Child development}
  1969\string: 969--1025.

\bibitem{page2014product}
Page T. Product attachment and replacement: implications for sustainable
  design. {\it International Journal of Sustainable Design} 2014\string;
  2(3)\string: 265--282.

\bibitem{Mugge}
Mugge R. {\it Product attachment}. PhD thesis. Delft University of Technology,
  2007.

\bibitem{mugge2008product}
Mugge R, Schoormans JP, Schifferstein HN. Product attachment-17: Design
  strategies to stimulate the emotional bonding to products.  2008.

\bibitem{ball1992role}
Ball AD, Tasaki LH. The role and measurement of attachment in consumer
  behavior. {\it Journal of Consumer Psychology} 1992\string; 1(2)\string:
  155--172.

\bibitem{kleine1995possession}
Kleinef SS, Kleine~III RE, Allen CT. How is a possession" me" or" not me"?
  Characterizing types and an antecedent of material possession attachment.
  {\it Journal of Consumer Research} 1995\string: 327--343.

\bibitem{mugge2009emotional}
Mugge R, Schoormans JP, Schifferstein HN. Emotional bonding with personalised
  products. {\it Journal of Engineering Design} 2009\string; 20(5)\string:
  467--476.

\bibitem{mugge2009development}
Mugge R, Govers PC, Schoormans JP. The development and testing of a product
  personality scale. {\it Design Studies} 2009\string; 30(3)\string: 287--302.

\bibitem{govers2005product}
Govers PC, Schoormans JP. Product personality and its influence on consumer
  preference. {\it Journal of Consumer Marketing} 2005\string; 22(4)\string:
  189--197.

\bibitem{sirgy1982self}
Sirgy MJ. Self-concept in consumer behavior: A critical review. {\it Journal of
  consumer research} 1982\string: 287--300.

\bibitem{meschtscherjakov2014mobile}
Meschtscherjakov A, Wilfinger D, Tscheligi M. Mobile attachment causes and
  consequences for emotional bonding with mobile phones. In:  {\it Proceedings
  of the 32nd annual ACM conference on Human factors in computing systems}ACM.
  ; 2014\string: 2317--2326.

\bibitem{champoux2016organizational}
Champoux JE. {\it Organizational behavior: Integrating individuals, groups, and
  organizations}.
\newblock Routledge .
\newblock 2016.

\bibitem{klein2012memory}
Klein SB, Nichols S. Memory and the sense of personal identity. {\it Mind}
  2012\string: 677--702.

\bibitem{csikszentmihalyi1981meaning}
Csikszentmihalyi M, Halton E. {\it The meaning of things: Domestic symbols and
  the self}.
\newblock Cambridge University Press .
\newblock 1981.

\bibitem{mugge2010product}
Mugge R, Schifferstein HN, Schoormans JP. Product attachment and satisfaction:
  understanding consumers' post-purchase behavior. {\it Journal of consumer
  Marketing} 2010\string; 27(3)\string: 271--282.

\bibitem{Stanford}
https://plato.stanford.edu/entries/hedonism/, [cited on-line: accessed
  24/01/2017].  2017.

\bibitem{jordan2002designing}
Jordan PW. {\it Designing pleasurable products: An introduction to the new
  human factors}.
\newblock CRC press .
\newblock 2002.

\bibitem{glanze1990mosby}
Glanze WD, Anderson K, Anderson LE, others . {\it Mosby's medical, nursing, and
  allied health dictionary}.
\newblock Mosby .
\newblock 1990.

\bibitem{callele2008balancing}
Callele D, Neufeld E, Schneider K. Balancing security requirements and
  emotional requirements in video games. In:  {\it International Requirements
  Engineering, 2008. RE'08. 16th IEEE}IEEE. ; 2008\string: 319--320.

\bibitem{lopez2014modelling}
Lopez-Lorca AA, Miller T, Pedell S, Sterling L, Curumsing MK. Modelling
  Emotional Requirements. 2014.

\bibitem{chapman2015emotionally}
Chapman J. {\it Emotionally durable design: objects, experiences and empathy}.
\newblock Routledge .
\newblock 2015.

\bibitem{schifferstein2004designing}
Schifferstein HN, Mugge R, Hekkert P. Designing consumer-product attachment.
  In:  {\it Design and Emotion}CRC Press. ; 2004\string: 378--83.

\bibitem{ingram1984designing}
Ingram J. Designing the spatial experience. {\it Design Studies} 1984\string;
  5(1)\string: 15--20.

\bibitem{marcus2015emotion}
Marcus A. The Emotion Commotion. In:  {\it HCI and User-Experience
  Design}Springer; 2015\string: 83--89.

\bibitem{bates1994role}
Bates J, others . The role of emotion in believable agents. {\it Communications
  of the ACM} 1994\string; 37(7)\string: 122--125.

\bibitem{salen2004rules}
Salen K, Zimmerman E. {\it Rules of play: Game design fundamentals}.
\newblock MIT press .
\newblock 2004.

\bibitem{ramos2005emotion}
Ramos I, Berry DM. Is emotion relevant to requirements engineering?. {\it
  Requirements Engineering} 2005\string; 10(3)\string: 238--242.

\bibitem{colomo2011using}
Colomo-Palacios R, Casado-Lumbreras C, Soto-Acosta P, Garc{\'\i}a-Crespo {\'A}.
  Using the affect grid to measure emotions in software requirements
  engineering. {\it Journal of Universal Computer Science} 2011\string;
  17(9)\string: 1281--1298.

\bibitem{thewvalue}
Thew S, Sutcliffe A. Value-based requirements engineering: method and
  experience. {\it Requirements Engineering} 2017\string: 1--22.

\bibitem{thew2008investigating}
Thew S, Sutcliffe A. Investigating the Role of'Soft issues' in the RE Process.
  In:  {\it International Requirements Engineering, 2008. RE'08. 16th
  IEEE}IEEE. ; 2008\string: 63--66.

\bibitem{mumford2013values}
Mumford E. {\it Values, technology and work}. 3.
\newblock Springer Science \& Business Media .
\newblock 2013.

\bibitem{sommerville1993integrating}
Sommerville I, Rodden T, Sawyer P, Bentley R, Twidale M. Integrating
  ethnography into the requirements engineering process. In:  {\it Requirements
  Engineering, 1993., Proceedings of IEEE International Symposium on}IEEE. ;
  1993\string: 165--173.

\bibitem{hughes1995presenting}
Hughes J, O'Brien J, Rodden T, Rouncefield M, Sommerville I. Presenting
  ethnography in the requirements process. In:  {\it Requirements Engineering,
  1995., Proceedings of the Second IEEE International Symposium on}IEEE. ;
  1995\string: 27--34.

\bibitem{viller1999social}
Viller S, Sommerville I. Social analysis in the requirements engineering
  process: from ethnography to method. In:  {\it Requirements Engineering,
  1999. Proceedings. IEEE International Symposium on}IEEE. ; 1999\string:
  6--13.

\bibitem{eric2009social}
Eric SY. Social Modeling and i*. In:  {\it Conceptual Modeling: Foundations and
  Applications}Springer; 2009\string: 99--121.

\bibitem{eric2010social}
Eric SY. Introduction to social modelling in RE. In:  Eric Y, Paolo G, Neil M,
  Mylopoulos J. \kern-2pt, eds. {\it Social modelling for requirements
  engineering}The MIT Press; 2010.

\bibitem{friedman1996value}
Friedman B. Value-sensitive design. {\it interactions} 1996\string;
  3(6)\string: 16--23.

\bibitem{friedman2002value}
Friedman B, Kahn P, Borning A. Value sensitive design: Theory and methods. {\it
  University of Washington technical report} 2002\string: 02--12.

\bibitem{cockton2009evolving}
Cockton G, Kirk D, Sellen A, Banks R. Evolving and augmenting worth mapping for
  family archives. In:  {\it Proceedings of the 23rd British HCI Group Annual
  Conference on People and Computers: Celebrating People and Technology}British
  Computer Society. ; 2009\string: 329--338.

\bibitem{gordijn2003value}
Gordijn J, Akkermans J. Value-based requirements engineering: exploring
  innovative e-commerce ideas. {\it Requirements engineering} 2003\string;
  8(2)\string: 114--134.

\bibitem{komssi2011integrating}
Komssi M, Kauppinen M, T{\"o}h{\"o}nen H, Lehtola L, Davis AM. Integrating
  analysis of customers' processes into roadmapping: The value-creation
  perspective. In:  {\it Requirements Engineering Conference (RE), 2011 19th
  IEEE International}IEEE. ; 2011\string: 57--66.

\bibitem{fuentes2010understanding}
Fuentes-Fern{\'a}ndez R, G{\'o}mez-Sanz JJ, Pav{\'o}n J. Understanding the
  human context in requirements elicitation. {\it Requirements engineering}
  2010\string; 15(3)\string: 267--283.

\bibitem{kleine2004integrative}
Kleine SS, Baker SM. An integrative review of material possession attachment.
  {\it Academy of marketing science review} 2004\string; 2004\string: 1.

\bibitem{winfree2017learning}
Winfree T, Goldacre P, Sherkat M, Graham P, Mendoza A, Miller T. Learning for
  Low Carbon Living: The Potential of Mobile Learning Applications for Built
  Environment Trades and Professionals in Australia. {\it Procedia Engineering}
  2017\string; 180\string: 1773--1783.

\bibitem{maclachlan2009exploring}
Maclachlan M, Harrison D, Wood B, others . Exploring the Reflective and
  Utilitarian Benefits of Product Attachment. In:  {\it DS 58-10: Proceedings
  of ICED 09, the 17th International Conference on Engineering Design, Vol. 10,
  Design Education and Lifelong Learning, Palo Alto, CA, USA, 24.-27.08.};
  2009.

\bibitem{maclachlan2009let}
Maclachlan M, Harrison D, Wood B, others . Let's get emotional: introducing
  undergraduate Product Design students to the concept of emotional design. In:
   {\it DS 59: Proceedings of E\&PDE 2009, the 11th Engineering and Product
  Design Education Conference-Creating a Better World, Brighton, UK,
  10.-11.09.}; 2009.

\bibitem{neuman2005social}
Neuman WL. {\it Social research methods: Quantitative and qualitative
  approaches}. 13.
\newblock Allyn and Bacon Boston .
\newblock 2005.

\bibitem{hatton2008choosing}
Hatton S. Choosing the right prioritisation method. In:  {\it Software
  Engineering, 2008. ASWEC 2008. 19th Australian Conference on}IEEE. ;
  2008\string: 517--526.

\bibitem{paetsch2003requirements}
Paetsch F, Eberlein A, Maurer F. Requirements engineering and agile software
  development. In:  {\it Enabling Technologies: Infrastructure for
  Collaborative Enterprises, 2003. WET ICE 2003. Proceedings. Twelfth IEEE
  International Workshops on}IEEE. ; 2003\string: 308--313.

\bibitem{moody2003method}
Moody DL. The method evaluation model: a theoretical model for validating
  information systems design methods. {\it ECIS 2003 proceedings} 2003\string:
  79.

\bibitem{pennotti2009evaluating}
Pennotti M, Turner R, Shull F. Evaluating the effectiveness of systems and
  software engineering methods, processes and tools for use in defense
  programs. In:  {\it Systems Conference, 2009 3rd Annual IEEE}IEEE. ;
  2009\string: 319--322.

\bibitem{zowghi2002three}
Zowghi D, Gervasi V. The Three Cs of requirements: consistency, completeness,
  and correctness. In:  {\it International Workshop on Requirements
  Engineering: Foundations for Software Quality, Essen, Germany: Essener
  Informatik Beitiage}; 2002\string: 155--164.

\bibitem{lee2014software}
Lee MC. Software Quality Factors and Software Quality Metrics to Enhance
  Software Quality Assurance. {\it British Journal of Applied Science \&
  Technology} 2014\string; 4(21)\string: 3069--3095.

\bibitem{pressman2005software}
Pressman RS. {\it Software engineering: a practitioner's approach}.
\newblock Palgrave Macmillan .
\newblock 2005.

\bibitem{trochim2001research}
Trochim WM, Donnelly JP. Research methods knowledge base.  2001.

\bibitem{leung2015validity}
Leung L. Validity, reliability, and generalizability in qualitative research.
  {\it Journal of family medicine and primary care} 2015\string; 4(3)\string:
  324.

\bibitem{davis1989perceived}
Davis FD. Perceived usefulness, perceived ease of use, and user acceptance of
  information technology. {\it MIS quarterly} 1989\string: 319--340.

\bibitem{fitzgerald1991validating}
Fitzgerald G. Validating new information systems techniques: a retrospective
  analysis. {\it Information systems research: Contemporary approaches and
  emergent traditions} 1991\string: 657--672.

\bibitem{mendoza2010learnability}
Mendoza A, Stern L, Carroll J. Learnability’as a positive influence on
  technology use. In:  {\it Electronic Proceedings of the 4th International
  Multi-Conference on Society, Cybernetics and Informatics, Retrieved from
  http://www. iiis. org/CDs2010/CD2010SCI/IMSCI\_2010/index. asp}; 2010.

\bibitem{mendoza2010software}
Mendoza A, Carroll J, Stern L. Software appropriation over time: from adoption
  to stabilization and beyond. {\it Australasian Journal of Information
  Systems} 2010\string; 16(2).

\bibitem{infoxchange}
http://www.infoxchange.net.au/Homeless-Assist-app-QandA, [cited on-line:
  accessed 12/01/2016].  2016.

\bibitem{sheskin2003handbook}
Sheskin DJ. {\it Handbook of parametric and nonparametric statistical
  procedures}.
\newblock crc Press .
\newblock 2003.

\bibitem{boudreau2001validation}
Boudreau MC, Gefen D, Straub DW. Validation in information systems research: a
  state-of-the-art assessment. {\it MIS quarterly} 2001\string: 1--16.

\bibitem{saeed2013computer}
Saeed K, Chaki R, Cortesi A, Wierzcho{\'n} S. {\it Computer Information Systems
  and Industrial Management: 12th IFIP TC 8 International Conference, CISIM
  2013, Krakow, Poland, September 25-27, 2013, Proceedings}. 8104.
\newblock Springer .
\newblock 2013.

\bibitem{paper1}
Sherkat M, Miller T, Mendoza A, Burrows R. Emotionalism within People-Oriented
  Software Design. {\it arXiv preprint arXiv:1810.12547v1} 2018.

\end{thebibliography}

\newpage
\appendix
\subsection*{Appendix A}
\label{append1}

\setcounter{table}{0}
\renewcommand{\thetable}{A.\arabic{table}}
\setcounter{figure}{0} \renewcommand{\thefigure}{A.\arabic{figure}}

\begin{figure}[ht]
\centering
\includegraphics [scale=0.6]{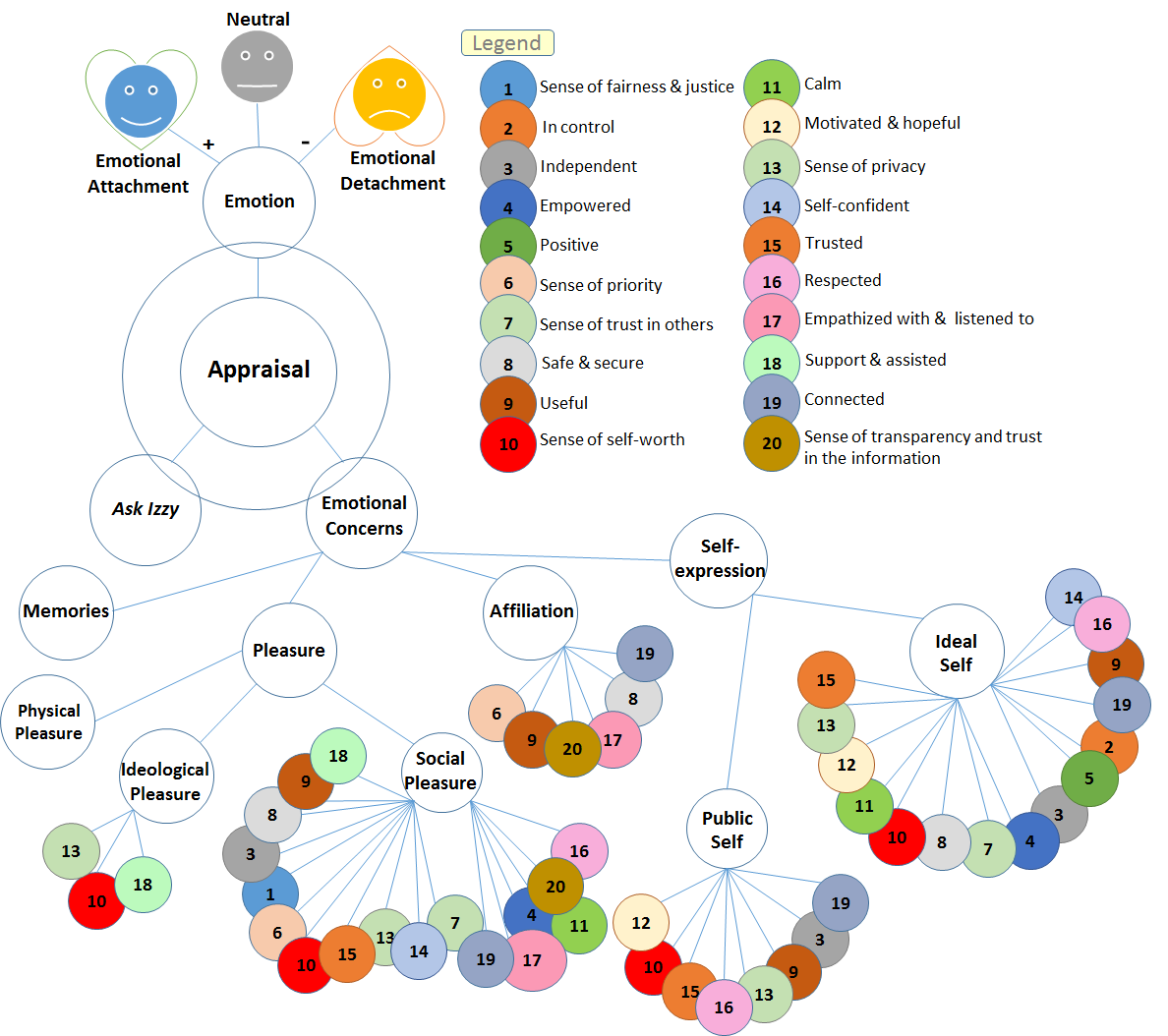}
\caption{\emph {Ask Izzy} Emotional Goals and Attachment Drivers}
\label{fig:conclusion}
\end{figure}

\end{document}